\documentclass[iop]{emulateapj}
\usepackage{amsmath}
\usepackage{amsbsy}
\usepackage{amssymb}
\usepackage{graphicx}
\usepackage{calrsfs}
\usepackage{natbib}
\usepackage[hypertex,breaklinks]{hyperref}

\begin{document}

\def\b{\boldsymbol}

\title{Synchrotron-to-curvature transition regime of  radiation of charged particles  in a dipole magnetic field}

\author{A.Yu.~Prosekin}
\affiliation{Max-Planck-Institut f\"ur Kernphysik,
Saupfercheckweg 1, D-69117 Heidelberg, Germany}
\email{Anton.Prosekin@mpi-hd.mpg.de}

\author{S.R.~Kelner}
\affiliation
{Max-Planck-Institut f\"ur Kernphysik,
Saupfercheckweg 1, D-69117 Heidelberg, Germany}
\affiliation{Research Nuclear University (MEPHI), Kashirskoe shosse 31,
115409 Moscow, Russia}
\email{Stanislav.Kelner@mpi-hd.mpg.de}

\author{F.A.~Aharonian}
\affiliation{Dublin Institute for Advanced Studies, 31 Fitzwilliam Place,
Dublin 2, Ireland}
\affiliation
{Max-Planck-Institut f\"ur Kernphysik,
Saupfercheckweg 1, D-69117 Heidelberg, Germany}
\email{Felix.Aharonian@mpi-hd.mpg.de}

\date{\today}

\begin{abstract}
The details of trajectories of charged particles become increasingly  important for  proper understanding of 
processes of  formation of  radiation  in  strong and curved magnetic fields. Because of  damping of the 
perpendicular component of motion,  the particle's pitch angle could be  decreased  by many orders of magnitude
leading to the change of the radiation regime  --      
from synchrotron to the curvature mode. To explore the character  of this  transition,    
we solve numerically the equations of motion of a test particle in a 
dipole  magnetic field, and calculate  the energy spectrum of magnetic bremsstrahlung self-consistently, i.e.
without {\it a priori} assumptions  on the radiation regime.  In this way we can trace the transitions  between
the synchrotron and curvature regimes,  as well  as study the third (intermediate or the so-called 
synchro-curvature)  regime.  We  briefly discuss  three interesting astrophysical scenarios,  the 
radiation of electrons in the pulsar magnetosphere in the polar  cap and outer gap models, as well as the  
radiation of ultrahigh energy protons in the magnetosphere  of a massive black hole, and demonstrate 
that in these models  
the  synchrotron, synchro-curvature and curvature regimes can be   realized with quite 
different  relative contributions to the total emission.
\end{abstract}

\keywords{ gamma rays: theory --- magnetic fields --- radiation mechanisms: nonthermal} 

\maketitle
\section{\label{sec:into} Introduction}
The efficiency of synchrotron radiation depends on the 
pitch angle between the magnetic field and the particle
velocity. The damping of the perpendicular motion in course of radiation reduces the pitch angle.
Typically, in moderate magnetic fields  the pitch angle changes slowly. 
Therefore, for  calculations  of synchrotron radiation, it is sufficient to  specify the initial pitch angle
distribution of particles. 

The situation is different  in  strong magnetic fields, namely when   the energy losses   
become  so  intensive  that on  fairly  short timescales 
the pitch angle  can be reduced by orders of magnitude.  
In this regard,  the  adequate theoretical treatment of  particle trajectories 
becomes a key issue  for  correct calculations of radiation properties.  
In a curved magnetic field the strong damping of the perpendicular
motion  causes  transition from   synchrotron to the curvature radiation regime of radiation. 
The solutions of  equations that describe the particle motion  with an inclusion of  
energy losses,  allow  us to  trace this  transition,  and thus to calculate
the magnetic bremsstrahlung without additional assumptions regarding the  radiation regime. 

In this paper, we study the case of  motion of a charged  particle  
in the dipole magnetic field, and  calculate self-consistently the radiation
spectrum taking into account the time-evolution of  particle's  energy,  coordinates, and  direction. 
We demonstrate that even  a small deviation of  particle's  initial direction from the magnetic
field line may have a strong impact on the character of radiation.  
Despite  the fast transition to the {\it final} (curvature) 
regime, the particle radiates away  the major fraction  of  its  energy  
in  the {\it initial} (synchrotron)  or {\it transitional}  (synchro-curvature)  regimes.  
Consequently,   the energy spectrum of  radiation may differ considerably  
from the spectrum of curvature radiation.

\section{\label{sec:gen} General comments}
In a strong magnetic field,  the motion of a charged particle perpendicular to the field is damped. The amount of
energy lost during this process  depends on the initial energy and the pitch angle. In the case of relativistic
motion along and perpendicular to the  magnetic field, $p_\parallel\gg mc$ and $p_\perp \gg mc$, particles 
can lose  a large fraction of their energy even if the initial pitch angle is small. Indeed, after the complete damping of the
perpendicular component of motion in a homogeneous magnetic field,  the
parallel momentum becomes \citep{Kelner2012}
\begin{equation}
p'_{\parallel}=p_{\parallel}\frac{mc}{\sqrt{p_{\perp}^{2}+m^2c^2}}.
\end{equation}
Then for small initial pitch angles $\alpha\ll 1$ and for the relativistic motion perpendicular to the magnetic field,
we obtain
that the final Lorenz factor  depends only on the initial pitch angle:
\begin{equation}
\gamma'\approx\frac{1}{\alpha} \, .
\end{equation}
It is convenient to rewrite this expression in the following form:
\begin{equation}
\gamma'\approx\frac{\gamma}{\gamma_{\perp}},
\end{equation}
where $\gamma_{\perp}=\gamma\sin\alpha$ is referred hereafter as perpendicular Lorenz factor. 
We see that at an  ultrarelativistic motion of particle, 
even a tiny deflection $\alpha=\gamma_{\perp}/\gamma$ from the magnetic
field line can cause large energy losses $\Delta E=E\left(1-1/\gamma_{\perp}\right)\approx E$.

The damping rate  depends on the strength of the  magnetic field.  The super-strong 
magnetic fields  surrounding of  compact astrophysical  objects,  such as   pulsars and black holes, 
cause  very fast damping of the perpendicular component of motion, and in this  way  force 
the particle to move along magnetic field lines. However, since the magnetic field lines are curved,  
the change of the field direction results in an appearance of the perpendicular velocity. The curvature of 
magnetic field lines and the presence of the perpendicular velocity lead to the curvature drift (an averaged motion
perpendicular to the magnetic field due to non-compensated differences in the trajectory of the periodic motion
arising from changes of the magnetic field direction).
Thus,  after losing most its  
perpendicular motion in a curved magnetic field,   the particle 
moves along the {\it drift} trajectory gyrating around it. 

The radiation spectrum is determined by the curvature of particle's  trajectory which is a superposition of 
the drift and  the gyration around it. As long as the
real trajectory is close to the magnetic field lines,   for  calculations of radiation spectra one can use 
the curvature of the magnetic field lines  instead of the trajectory curvature. The difference between 
the magnetic field curvature and the drift trajectory curvature can be neglected  until the curvature drift
velocity is small. In case of the magnetic field of a  long straight wire,  the curvature of the drift
trajectory is
\begin{equation}
K_D=\frac{1}{r_0(1+\beta_D^2)}\approx\frac{1-\beta_D^2}{r_0},
\end{equation} 
where $r_0$ is the curvature radius of the magnetic field, $\beta_D$ is the drift velocity in the
units of the speed of light expressed as \citep{Alfven1963}
\begin{equation}{\label{eq:driftv}}
\beta_D=\frac{v_D}{c}=\frac{c}{\Omega r_0}=\frac{mc^2\gamma}{eB r_0},
\end{equation}
where $\Omega=eB/mc\gamma$ is the gyration frequency, and $B$ is the strength of the magnetic field.
The gyration itself introduces a much larger difference if the velocity perpendicular to the drift
trajectory $\beta_\perp$ is of the order of the drift velocity \citep{Kelner2012}
\begin{equation}\label{eq:curv}
K=\frac{1}{r_0}\sqrt{\displaystyle{1-2\frac{\beta_\perp}{\beta_D}\cos\tau+\frac{\beta_\perp^2}{\beta_D^2}}}.
\end{equation} 

Note that  the particle in the magnetic field of a  long straight wire can move strictly along
the drift trajectory without gyration ($\beta_\perp=0$) and therefore 
with minimum possible energy losses. Because of the 
possibility of such motion,  and treating   the motion with gyration as  a perturbation,
\cite{Kelner2012}  has called the  drift trajectory  as a  ``smooth trajectory''. 
Having the potential minimum on the drift trajectory, the particle with any initial pitch angle   
due to energy losses will asymptotically reach the motion strictly along the drift trajectory. 
In the case of an arbitrary magnetic field, when
the curvature is variable, it is not possible to make a definite statement, except that the particle tries to 
reach the potential minimum according to the local values of the magnetic field. If the curvature changes
slowly,  the particle motion  could become  very close to the drift trajectory ($\beta_\perp\approx 0$), but, 
because of gyration, never strictly approaches  it ($\beta_\perp=0$)  as in the case of magnetic field of 
an  infinitely long straight wire.

The  energy loss rate  and the  radiation spectrum behave   differently from the case of the curvature
radiation  when  the velocity component perpendicular to the drift trajectory $\beta_\perp$  is comparable or  larger  than the drift
velocity $\beta_D$. Because of gyration,  the radiation  is expected to 
be  different from the pure curvature radiation, even when  initially the particle moves  
strictly along the magnetic field line ($\beta_\perp=\beta_D$). 
The nominal curvature radiation is  generally treated  as  the  same synchrotron 
radiation  \citep{Landau2} with a spectral maximum at the characteristic  frequency $\omega_*$ 
corresponding to the curvature radius of the magnetic field line instead of the curvature radius of the real trajectory.
The substitution of the real trajectory curvature affects the radiation spectrum, namely  shifts the 
maximum to $\approx(1+\beta_\perp/\beta_D)\omega_*$.  Thus, the difference
from the curvature radiation  becomes significant  when  $\beta_\perp/\beta_D\gg 1$.  If  the pitch angle
is much larger than the angle between the magnetic field line and the drift trajectory,  one can
neglect it  and consider $\beta_\perp$ as the velocity perpendicular to magnetic field
line (although $\beta_\perp$ is the velocity perpendicular to the drift trajectory). It implies the  dominance
of  the ratio $\beta_\perp/\beta_D$ in the expression $1+\beta_\perp/\beta_D$, i.e.  the 
radiation deviates significantly from the  curvature radiation when the velocity perpendicular to the
magnetic field is greater than the drift velocity.

It is interesting to  study the possibility of  of pitch angles greater than $\beta_D$.
This question is specific and its  answer depends on the acceleration mechanism, but here we would like to discuss
very general points. The pitch angle of the particle accelerated by the electric field in the presence of
magnetic field depends on the relation between strengths of the fields and the angle between them. It is worth
to mention the result of acceleration in the crossed homogeneous fields. In this case one can find the
reference frame where the fields are parallel. The particle in this reference frame is infinitely accelerated
along the parallel fields and  lose the perpendicular component of the motion. Taking the motion along the field
and transforming the velocity back to the laboratory reference frame, the pitch angle can be expressed as
\begin{equation}
\sin \alpha=\frac{\rho \sin
\theta}{\sqrt{\frac{1}{2}\left(\sqrt{(1+\rho^2)^2-4\rho^2\sin^2\theta}+(1+\rho^2)\right)}},
\end{equation}
where $\rho=E/B$ is the ratio of the  electric and magnetic fields, and 
 $\theta$ is the angle between them.  Assuming that
typically the electric field is smaller than magnetic field,  we obtain 
\begin{equation}
\sin \alpha\approx\rho \sin\theta
\end{equation}
which states that the pitch angle equals the drift velocity  due to electric drift (in the units of the speed of
light).

In a more general case,  a drift   due to the electric field should appear as well. Then we arrive at  a quite 
general conclusion that the radiation of the accelerated particle could be considerably different from the
curvature radiation if the electrical drift exceeds the curvature drift
\begin{equation}
\rho\sin\theta \gg \beta_D.
\end{equation}

The so-called outer gap model of pulsars gives an example of existence of a perpendicular component of the electric field. In the acceleration gaps, the electric field could not be parallel to the magnetic field everywhere, in particular
close to the border of the gap the perpendicular component of the electric field is  increased  \citep{Cheng1986}.
In these regions  the radiation can be  different from the curvature radiation.


\section{Local trajectory}
The  accurate calculation of the radiation spectrum requires  detailed knowledge of  the trajectory of charged particle. The latter  obtained in the drift approximation is not suitable for this purpose since 
the fast gyrations  are erased in the course of  averaging.  Fortunately, 
under the assumption that the gyroradius is small,  it is possible to obtain a local solution which  takes  into account  fast gyrations for the motion in an arbitrary curved magnetic field  \cite{Trubnikov2000}.  This approach is equivalent to the consideration of particle  motion in the magnetic field which is
constant along binormal and has a constant curvature and zero torsion. The magnetic lines of this field
present circles with centres  lying on a straight line. The magnetic field of this structure is created by the
current of an  infinitely long straight wire. It allows us to consider the solution obtained in 
\cite{Kelner2012} for such field as a local one. This consideration is also
possible because a small part of the curved magnetic field line can always be approximated by the  arc of a
circle with the radius equal to the curvature radius of the line. 

We consider the particle motion in the local coordinate system:
\begin{equation}
\b r=s \b h+x \b n+y \b k,
\end{equation} 
where $\b h=\b B/B$ is the unit vector in the direction of the magnetic field, 
$\b k=\b h\times \b n$ is the binormal vector, 
$\b n=r_0(\b h\nabla)\b h$ is the normal vector with  $r_0$ as  the curvature radius. Then, in accordance with \cite{Kelner2012},  the local 
velocity can be written as
\begin{equation}
\b v=v_s \b h+v_x \b n+v_y \b k,
\end{equation}  
where
\begin{eqnarray}\label{eq:Vel}
\begin{aligned}
v_{s}&=c\beta_{\parallel}\left(1+\frac{\beta_{\perp}\beta_D}{\beta_{\parallel}^2}\cos\tau \right),\\
v_{x}&=c\beta_{\perp}\sin\tau,\quad
v_{y}=c\beta_D-c\beta_{\perp}\cos\tau.
\end{aligned}
\end{eqnarray} 
Here $\beta_{\parallel}$ is the component of the velocity  along magnetic field line,
$\beta_{\perp}$ is the component of the velocity perpendicular to the drift trajectory, $\beta_{D}$ is the
drift velocity defined in Eq.~(\ref{eq:driftv}) (all velocities are in units of $c$), $\tau=\Omega\, t$ is the time in the units of the gyration
period $1/\Omega$.

The expressions in Eq.~(\ref{eq:Vel}) describe the motion along the magnetic field line with gyration around it and the drift
in the direction of the binormal to the magnetic field line. The solution formally includes also the case of 
strict motion along the drift trajectory without gyration ($\beta_{\perp}=0$). Such a  situation can occur only locally.
Since the curvature of the magnetic field lines changes, the gyration does not disappear completely although it
could be very small compared
to the drift. Eq.~(\ref{eq:Vel}) is  correct if the radius of gyration is much smaller than the curvature radius.
This is equivalent to the conditions $\beta_{\perp}\ll \beta_{\parallel}$ and $\beta_{D}\ll \beta_{\parallel}$.
Moreover, it allows us to neglect the drift gradient which, otherwise, would lead (in the case of the vacuum magnetic
field $\nabla\times\b B=0$) to the drift velocity
\begin{equation}
v_{D}=\frac{v_{\parallel}^2+\frac{v_{\perp}^2}{2}}{r_0\Omega}.
\end{equation}

The expression for the acceleration in the magnetic field
\begin{equation}
\b a=\frac{e}{mc\gamma}(\b v\times\b B),
\end{equation}
and Eq.~(\ref{eq:Vel}) allow us to find the absolute value of the acceleration
\begin{equation}
a=a_0\sqrt{1-2\eta\cos\tau+\eta^2},
\end{equation}
where $a_0=c^2\beta_{\parallel}^2/r_0$ is the acceleration due to curvature of the magnetic field, 
$\eta=\beta_{\perp}/\beta_D$ is the ratio  between the velocity perpendicular to the drift trajectory and
the drift velocity. It is convenient to introduce, as suggested in \cite{Kelner2012}, 
a  parameter which shows the
difference between the total
acceleration and the acceleration induced by the magnetic field curvature
\begin{equation}\label{eq:qloc}
q(\eta,\tau)=\frac{a}{a_0}=\sqrt{1-2\eta\cos\tau+\eta^2}.
\end{equation} 
Because of the simple relation between the acceleration and the curvature radius of the trajectory in the
ultrarelativistic case $a=c^2/R_c$, the $q$-parameter indicates also the difference between the trajectory
curvature and the curvature of the magnetic field lines (compare with Eq.~(\ref{eq:curv}))
\begin{equation}\label{eq:rc}
R_c=\frac{r_0}{q}.
\end{equation}
As  discussed in Sec. 2,  we will  neglect the difference between the curvature of the drift trajectory and the curvature of the magnetic field line.

The solution given by Eq.~(\ref{eq:Vel}) can be applied locally if the parameters of the motion such as the energy of
the particle, the curvature and the strength of the magnetic field are changed  slowly. In this case,  the
parameters of the solution $\Omega$, $\beta_{\perp}$ and $\beta_D$ change slowly as well. The $q$-parameter
varies in the limits $|1-\eta|\leqslant q \leqslant 1+\eta$. If $\eta\leqslant 1$,  then $q$ appears within the band of
width $2\eta$ around $q=1$. If $\eta>1$,  then $q$ varies  in the band of width $2$ around $q=\eta$. So
over the period of gyration,  $q$ could not change greater than $2$. Naturally, if the parameters of the motion
change considerably over the gyration period, the behaviour of $q$-parameter would be different.

\section {Radiation spectrum}
The spectral power density of the synchrotron radiation is defined by the instantaneous curvature
radius of the particle trajectory \citep{Schwinger1949}. For  the curvature radius given by Eq.~(\ref{eq:rc})
the spectral power density of radiation of the particle moving in the curved magnetic field 
can be written in the form
\begin{equation}\label{eq:InRad}
P(\omega,t)=\frac{\sqrt{3}e^2}{2\pi}\frac{\gamma}{r_0}G\left( \frac{\omega}{\omega_*} \right), 
\end{equation}
where
\begin{equation} \label{eq:Grad}
G\left( \frac{\omega}{\omega_*} \right)=q F\left(\frac{\omega} {\omega_{*} q} \right).
\end{equation}
Here $\omega_*=3c\gamma^3/(2r_0)$ is the characteristic  frequency of the curvature radiation and
\begin{equation}\label{eq:fx}
F(x)=x\int_x^{\infty}K_{5/3}(u) du
\end{equation}
is the emissivity function of  synchrotron radiation.
The  $q$-parameter  defined  as the 
ratio between the total acceleration and the acceleration induced by the
curvature of the magnetic field,  can be  expressed as
\begin{equation}\label{eq:qgen}
q=\frac{a}{a_0}=\frac{evB\sin\alpha}{mc\gamma}/\frac{v_{\parallel}^2}{r_0}\approx \frac{\sin\alpha}{\beta_D},
\end{equation}
where $\alpha$ is the pitch angle, $v$ is the particle velocity, $v_{\parallel}$ is the velocity along the
magnetic field. If the energy and the magnetic field are  changed  slowly,  the local representation given by 
Eq.~(\ref{eq:qloc}) can be used. Then the spectrum averaged over the period of the gyration is
determined by the function
\begin{equation} \label{eq:Grad1}
\left\langle  G\left( \frac{\omega}{\omega_*} \right)\right\rangle=\frac{1}{\pi}\int_{0}^{\pi}q(\eta,\tau)
F\left(\frac{\omega}{\omega_{*} q(\eta,\tau)} \right) d\tau.
\end{equation}
In all cases under consideration the form of the spectrum is defined by the same function $F(x)$. The relevant 
parameters change only the position of the maximum and the intensity.
However, the function $F(x)$ should be changed to its quantum analogue  (see Eq.~\ref{eq:fq})  if
the parameter $\chi=B\gamma \sin\alpha/B_{cr}$ is of the order of unity and higher, where
$B_{cr}=2m^2c^3/3e\hbar\approx 2.94\cdot 10^{13}$~G \citep{Landau4}. 
Note that at such conditions,  the energy of the produced photon
is close to the energy of the radiating electron. The electron-positron pair production by a gamma-ray 
photon in the strong magnetic field occurs when the parameter $\chi\gtrsim 1$, where $\gamma$ is the photon energy 
in the units of the electron  rest mass, and $\alpha$ is the angle between the photon and magnetic field. In the curved magnetic field the angle between photon and magnetic field could become sufficiently large for production of electron-positron pairs. This could lead, 
provided that the optical depth is large,  to development of electromagnetic cascade and formation of 
radiation which is considerably different from the initial one.

The substitution of the $q$-parameter in the general form of Eq.~(\ref{eq:qgen}) to
Eq.~(\ref{eq:Grad}) results in the standard spectral power density of the synchrotron radiation. Thus the 
$q$-parameter expresses the difference between the conventional curvature radiation (when the curvature
of the trajectory is accepted to be equal to the curvature of the magnetic field line) and the actual radiation
which is the small pitch angle synchrotron radiation. One can see that there is no well-defined boundary
between the curvature and the synchrotron radiation. It seems natural to define the magnetic bremsstrahlung 
as {\it curvature radiation}  when the main contribution is  introduced by the curvature of the magnetic field line. This case corresponds to $\eta \ll 1$; see Eqs.~(\ref{eq:curv}) and (\ref{eq:qloc}). 
Then the {\it synchrotron radiation}  occurs when  the curvature of the trajectory
induced by the strength of the magnetic field provides the main contribution to the radiation. This corresponds to
$q\approx\eta \gg 1$.  Finally, at $\eta\sim 1$ both the strength and the curvature of the magnetic field play equal
role in production of emission which we will call  {\it synchro-curvature} radiation. 

The limits of applicability of Eq.~(\ref{eq:InRad}) defining the energy spectra of  the synchrotron and the
curvature radiation originate from the approach of its derivation proposed by Swinger (\cite{Schwinger1949}). The
radiation of the ultrarelativistic  particle is concentrated in a  narrow cone with the opening angle $\sim
1/\gamma$ and is therefore collected while the angle between the velocity and the direction of the observation
is of the order of the same of $\sim 1/\gamma$. The Swinger method is based on expansion of the trajectory in the 
small time interval. During this time interval,  the entire observable radiation should be collected. It means that the 
approach works if the particle velocity changes the direction at an  angle larger than $1/\gamma$ while the expansion
is valid. The analysis of the local trajectory given by Eq.~(\ref{eq:Vel}) in the curved magnetic field
gives the following  limits of applicability:
\begin{eqnarray}\label{eq:applim}
\begin{aligned}
\beta_{\perp}&\gg \frac{1}{\gamma},\quad \text{if}\quad 
\beta_{\perp}\gtrsim \beta_D,\\
\sqrt{\frac{\beta_D^3}{\beta_D^3+\beta_{\perp}}}&\gg \frac{1}{\gamma}, \quad
\text{if} \quad \beta_{\perp}\lesssim\beta_D.
\end{aligned}
\end{eqnarray}
The first condition corresponds to the case of synchrotron radiation and states that the perpendicular
motion should be relativistic as it is expected from the consideration of radiation in the homogeneous
magnetic field. The second condition corresponds to the situation for the curvature radiation of a particle 
with a small perpendicular momentum. Small gyrations almost do not influence on the applicability
of Eq.~(\ref{eq:InRad}) and in the limit $\beta_{\perp}=0$ the condition simply states that the motion
should be relativistic.

\section{Numerical implementation}
The analytical approach described above allows us to study the local properties of the particle trajectory 
and the radiation in the curved magnetic field. To solve the problem in the general case, taking into account the energy
losses of particles, we  performed numerical integration  of the equations of motion. 
The radiation properties have been studied in the dipole magnetic
field which seems to be  a quite  good approximation for the strong magnetic fields 
in compact astrophysical objects. The dipole magnetic
field has two distinct  features  to be taken into account. The first one is the fast decrease of the
strength with the distance  $\sim 1/r^3$ with a strong impact on  the radiation intensity. The second one is the
significant variation of the curvature  with a change of the polar angle $\theta$ from the dipole axis $\propto
\sin\theta/r$,  so the radiation spectra in the vicinity of the pole and the equator should be  different.

For numerical calculations we use the equations of motion in the ultrarelativistic limit. In this case
the radiation reaction force is opposite to the velocity which changes only its direction. The
equation of motion can be written in the form
\begin{equation}\label{eq:moteq}
mc\frac{d}{dt}(\gamma\b \beta)=e(\b \beta\times \b B)-|\b f|\b \beta,
\end{equation}
where $\b \beta$ is the velocity in units of $c$ with $|\b \beta|\approx 1$, $\b B$ is the magnetic field,
and $\b f$ is the radiation reaction force \citep{Landau2}. Taking the
scalar product of Eq.~(\ref{eq:moteq}) with velocity $\b \beta$ we obtain the differential
equation for the energy losses. This equation allows cancellation of the radiation reaction force with the
Lorenz factor time derivative in Eq.~(\ref{eq:moteq}).
Finally, the equation of motion has  the same form as for the consideration without energy losses,
where the energy enters as a parameter. For the sake of convenience of the numerical treatment and
comprehension of the structure of the system of equations, these
equations are written in the dimensionless
form:
\begin{align}
\frac{d\b r'}{d \tau}&=u_1 \b \beta\,, \\
\frac{d\b \beta}{d\tau}&=\xi(\b \beta \times\b b)\,, \\
\frac{d\xi}{d\tau}&=u_2(\b \beta \times\b b)^2\,\label{eq:LossEn}.
\end{align}
The system of equations depends on two dimensionless parameters
\begin{equation}
u_1=\frac{mc^2\gamma_0}{eB_0R_0} \quad  {\rm and} \quad
u_2=\frac{2}{3}\frac{e^3B_0}{m^2c^4}\gamma_0^2,
\end{equation}
where $\gamma_0$ is the initial Lorenz factor, $R_0$ is the characteristic distance to the radiating region from the
dipole, $B_0=B_*
\left(R_*/R_0\right)^3$ is the characteristic magnetic field with $R_*$ and $B_*$ being the  source radius and
the magnetic field at the pole of the dipole, $m$ is a particle mass, $c$ is the speed of light. Here we have
introduced the following dimensionless variables: $\b r'$ is the coordinate in the character units of length
$R_0$,  $\b \beta$ is the velocity of the particle in the units of $c$, $\xi=\gamma_0/\gamma$ is the ratio of
the initial and current value of the Lorenz factor, $\tau=t\,eB_0/mc\gamma_0$ is the characteristic  time in 
the units of the initial gyration period,  $\b b$ is the dimensionless dipole magnetic field which is expressed
as
\begin{equation}
\b b=\frac{3\b n(\b n \b \mu)-\b \mu}{2r'^3},
\end{equation} 
where $\b \mu$ is the unit vector in the direction of the dipole axes, $\b n=\b r'/r'$ is the unit vector
to the particle position.

It should be noted that  depending on the specific  conditions  characterizing an astrophysical source, 
the parameters $u_1$ and $u_2$ may differ by many orders of magnitude. For instance, for typical
parameters of the so-called {\it polar cap}  model of pulsar magnetosphere  $B_0=10^{12}$ G,
$R_0=10^6$ cm, and $\gamma_0=10^8$,  we have $u_1\approx 1.7\cdot 10^{-7}$ and $\qquad u_2\approx 1.1\cdot 10^{12}$.
Thus the problem is non-stiff, therefore  for  integration of  this system of differential equations 
the implicit Rosenbrock method has been used.

The calculations of the trajectory have been carried out for different initial conditions. The initial
position determined by the radius $R_0$ and the polar angle $\theta_0$ relative to the magnetic dipole axis
defines the typical environment parameters for the models under consideration. The radiation spectrum
has been studied for different initial pitch angles $\alpha_0$ and initial Lorenz factors $\gamma_0$.
The detailed knowledge of the trajectory and the energy allows us to find $q$-parameter from Eq.(\ref{eq:qgen})
and the radiation spectrum given by Eq.~(\ref{eq:InRad}) at any moment of time. The spectra integrated over the 
time, the so called  {\it cumulative spectrum},  have been obtained  for any initial condition under consideration.

Finally, in the case of a very  strong magnetic field and/or very  large Lorenz factor,  the particle may radiate 
in the quantum regime. More specifically,  when  the parameter $\chi=B\gamma \sin\alpha/B_{cr}$ 
becomes of the order of unity or larger,  Eq.~(\ref{eq:LossEn}) 
should be replaced by its  quantum analogue (see Appendix)
\begin{equation}
\frac{d\xi}{d\tau}=\frac{u_2(\b \beta \times\b b)^2}{\left(1+u_3
((\b \beta \times\b b)/\xi)^{2/3} \right)^2 },
\end{equation}
where $u_3=1.07\cdot 10^{-9}(B_0\gamma_0)^{2/3}$ and $B_{cr}=2m^2c^3/3e\hbar\approx 2.94\cdot 10^{13}$ G.

\section{Astrophysical implications}
In this section we explore   possible realizations of the  synchrotron and curvature regimes, as well as 
transitions between these two modes  of radiation (the synchro-curvature regime) in the context of three specific astrophysical scenarios.  Namely below we discuss the radiation of electrons and positrons in the pulsar magnetosphere for the outer gap and polar cap models,  and the radiation of protons at their  
acceleration in the vicinity of a supermassive  black hole.

\subsection{Outer Gap}
The high energy gamma radiation from pulsars is widely believed to be produced in the outer gap of the
pulsar magnetosphere \citep{Cheng1986,Takata2004,Hirotani2008}. 
Here we present the results of our calculations of radiation of electrons (positrons) 
in the dipole magnetic field at the location of the outer gap. The
position of the outer gap was assumed along the last open field line which is 
is determined by the inner boundary located at
the null surface where the Goldreich-Julian charge density is zero, and by the outer boundary taken on the surface 
of the light cylinder. We adopt the parameters of the Crab pulsar: the radius of the star  
$R_*=10^6$ cm, the rotation period  $P=33.5$ ms, and the magnetic field at the pole $B_*=10^{12}$ G.
We consider non-aligned pulsar with an  angle of $\pi/4$  between the rotational axis and the magnetic dipole
axis. The particle initial position is set  at the distance $r_{init}=0.5R_{lc}$ from the null surface along
the last open field line, where $R_{lc}$ is the radius of the light cylinder. During the numerical calculations, 
the particle is followed up to the intersection with the light cylinder. For the Crab parameters these conditions
correspond to $\theta_{0}=51.5^{\circ}$, $R_{0}=1.2\cdot 10^8$ cm, $B_{0}=5.6\cdot 10^{5}$ G, and
$R_{lc}=1.6\cdot 10^{8}$ cm, where $\theta_{0}$ is the polar angle relative to the magnetic dipole axis.

\begin{figure}
 \begin{center}
  \includegraphics[width=0.5\textwidth]{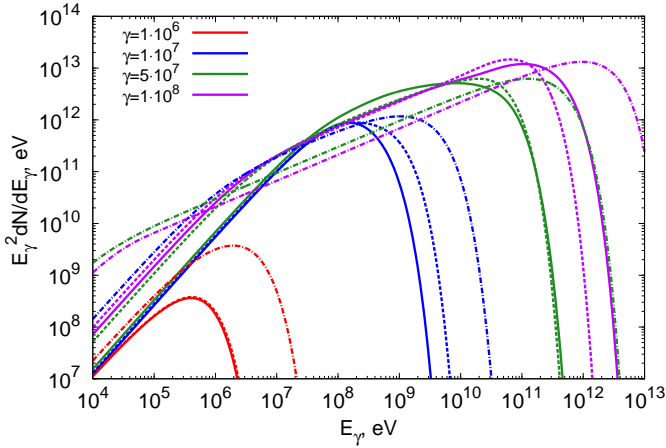}
  \caption{\label{fig:OGrad} The cumulative (integrated
  along trajectory) radiation spectra of electrons calculated for the 
  outer gap model of the pulsar magnetosphere. The curves are obtained  
  for different initial Lorenz factors  of electrons  $\gamma=10^6, 10^7, 5\times 10^7, 10^8$, and
  for different initial directions relative to the magnetic field lines:  along the drift trajectory (solid lines), 
  along the magnetic field line (dashed lines), and  at  pitch angle $\alpha=10\beta_D$ in the meridional plane opposite to the normal vector of magnetic
  field lines (dashed-dotted lines).}
 \end{center}
\end{figure}

\begin{figure}
\begin{center}$
\begin{array}{cc}
\includegraphics[width=0.25\textwidth]{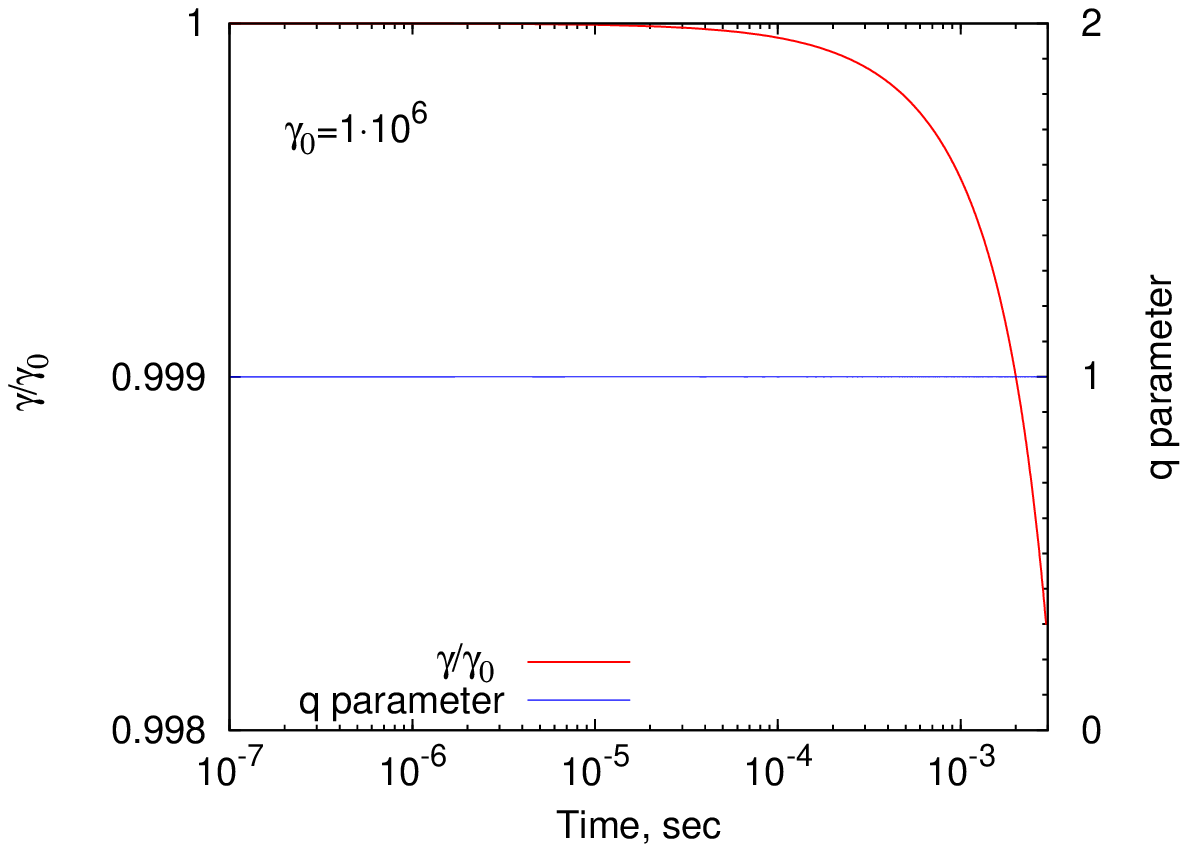} &
\includegraphics[width=0.25\textwidth]{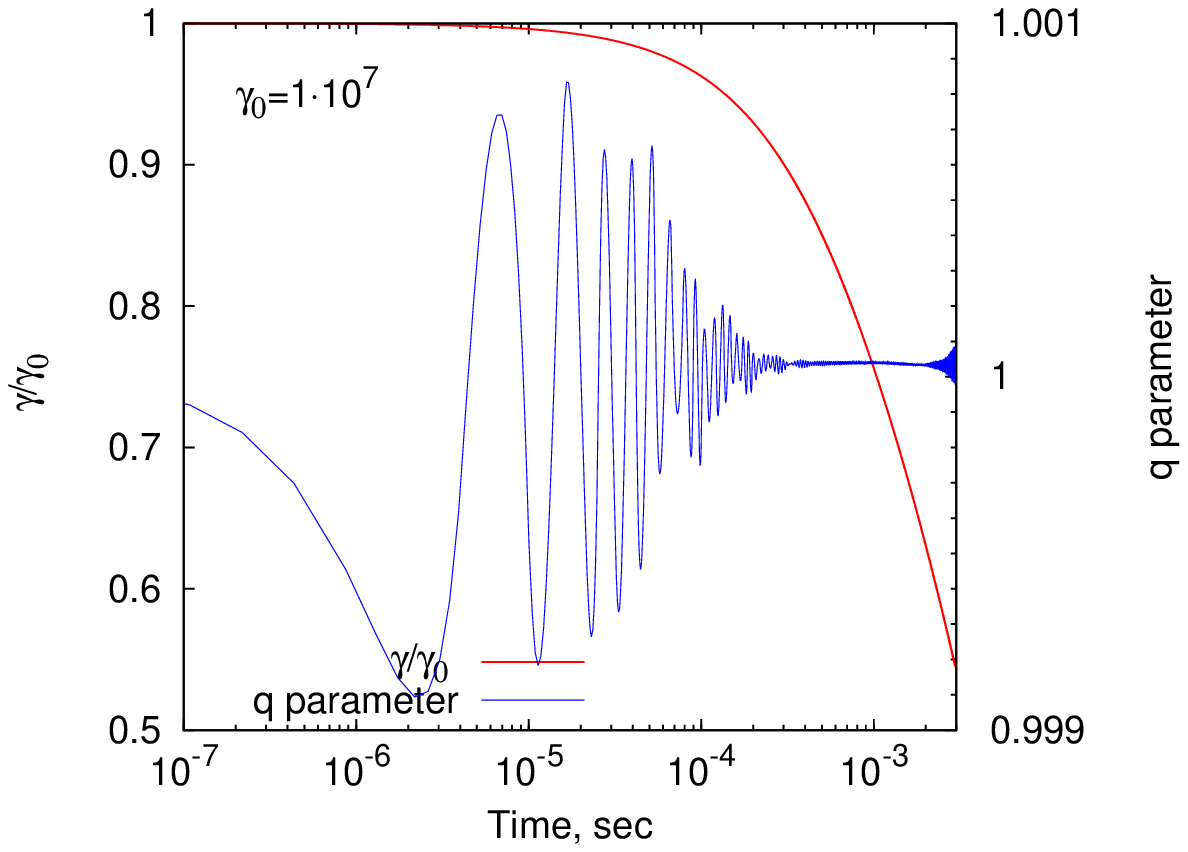} \\
\includegraphics[width=0.25\textwidth]{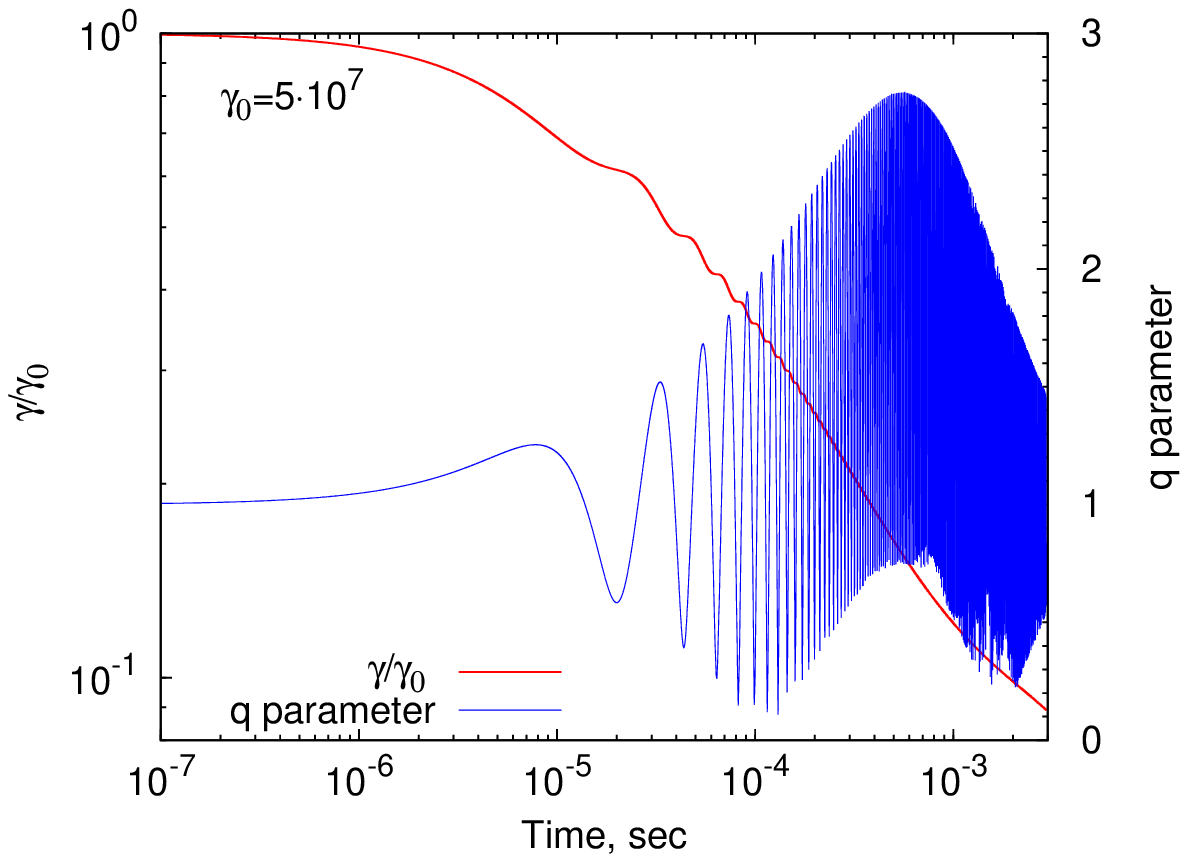} &
\includegraphics[width=0.25\textwidth]{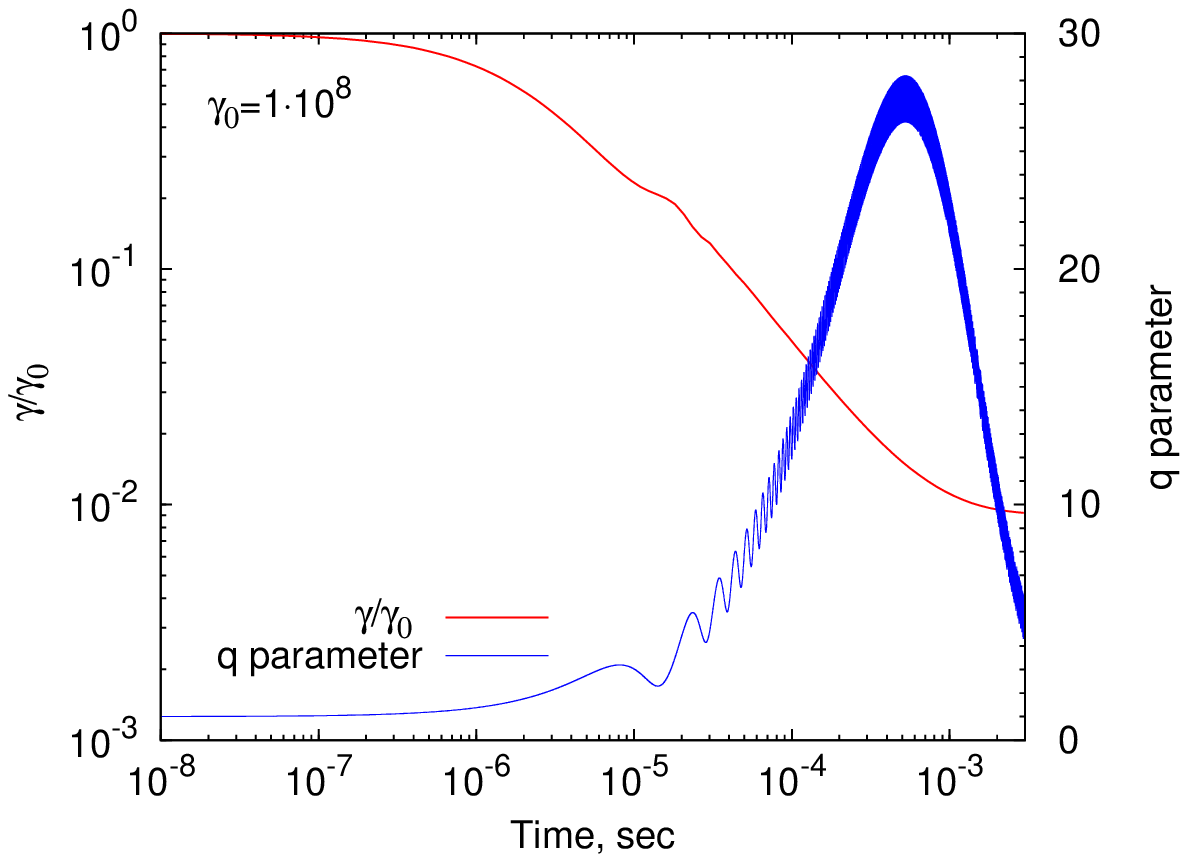}
\end{array}$
\end{center}
\caption{\label{fig:OGradComD}
Time evolution of the   $q$-parameter and the electron
 Lorenz factor in the outer gap model (complementary to Fig.~\ref{fig:OGrad}). Four panels correspond to 
 the initial Lorenz factor of  electrons  $\gamma=10^6, 10^7, 5\times 10^7, 10^8$
 and their initial direction  along the drift trajectory.}
\end{figure}

\begin{figure}
\begin{center}$
\begin{array}{cc}
\includegraphics[width=0.25\textwidth]{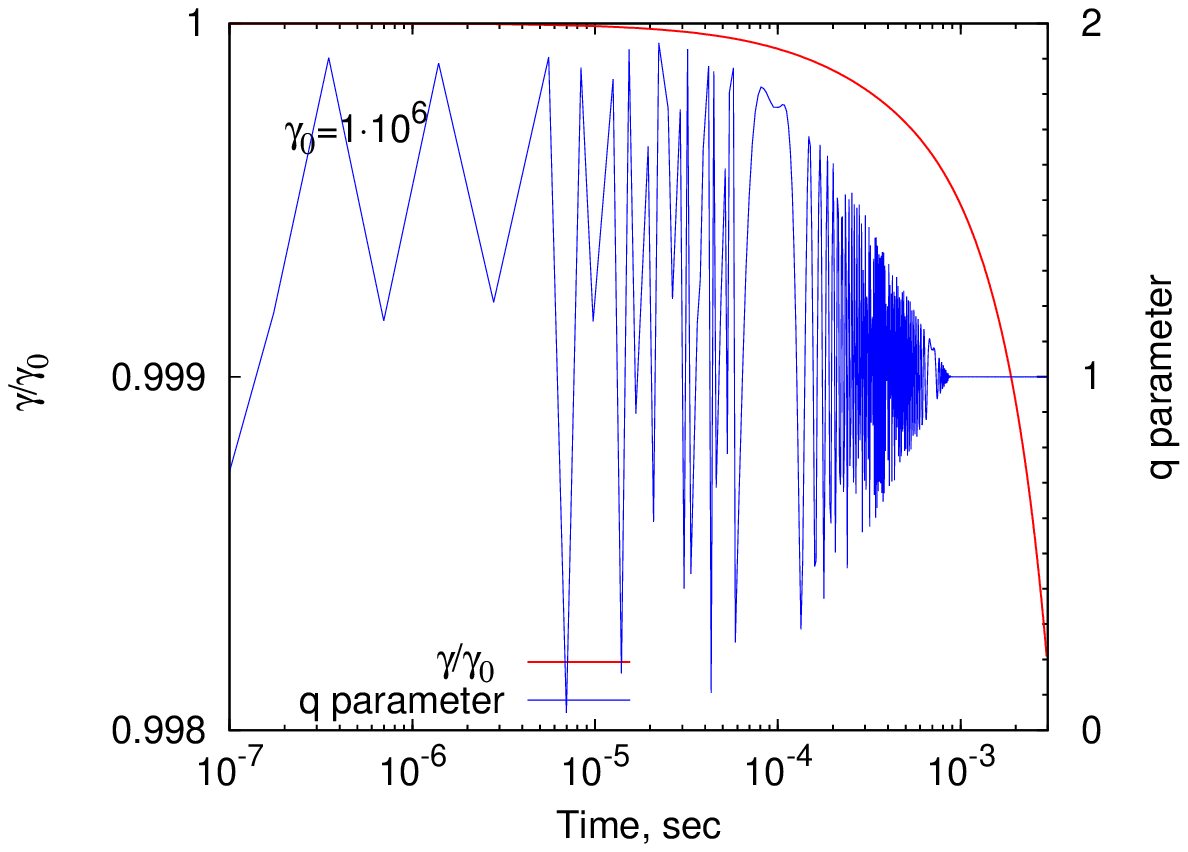} &
\includegraphics[width=0.25\textwidth]{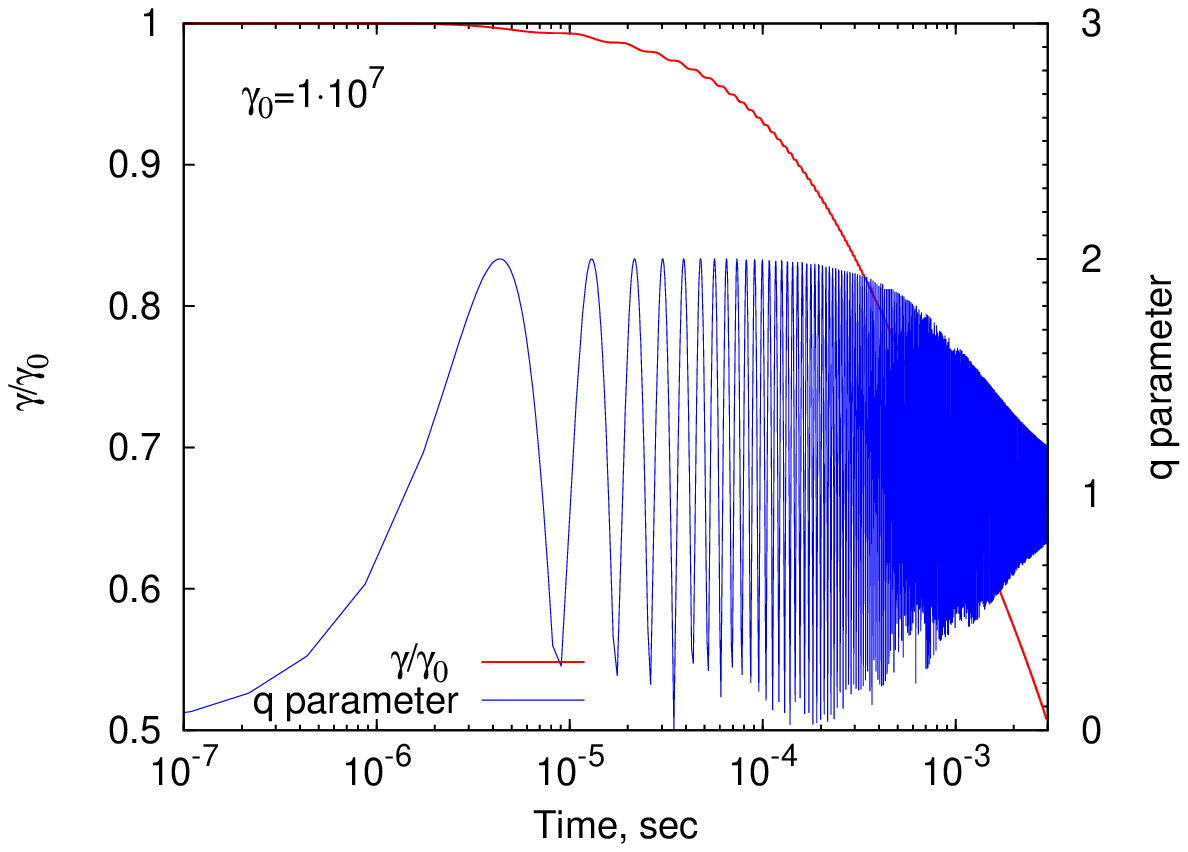} \\
\includegraphics[width=0.25\textwidth]{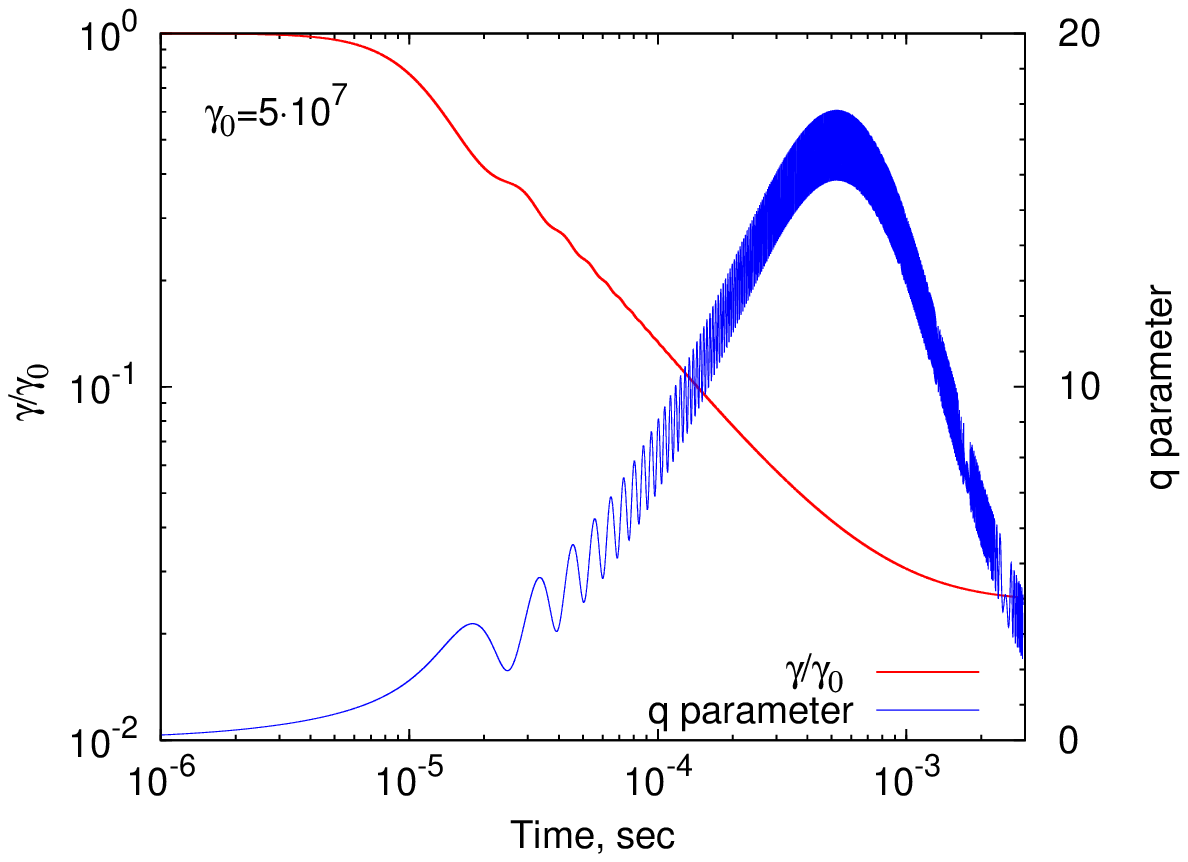} &
\includegraphics[width=0.25\textwidth]{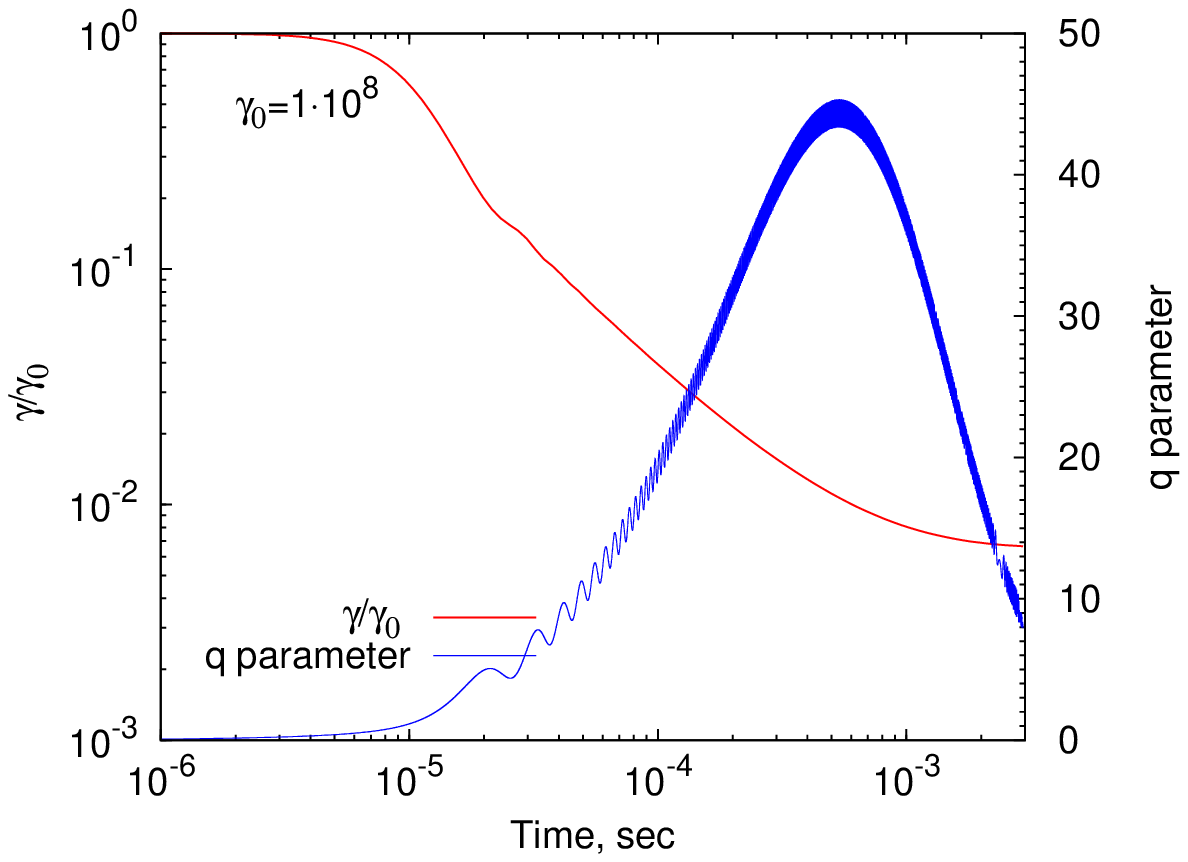}
\end{array}$
\end{center}
\caption{\label{fig:OGradComB}
The same as in Fig.\ref{fig:OGradComD}, but for the initial direction of electrons along the magnetic field line.}
\end{figure}

\begin{figure}
\begin{center}$
\begin{array}{cc}
\includegraphics[width=0.25\textwidth]{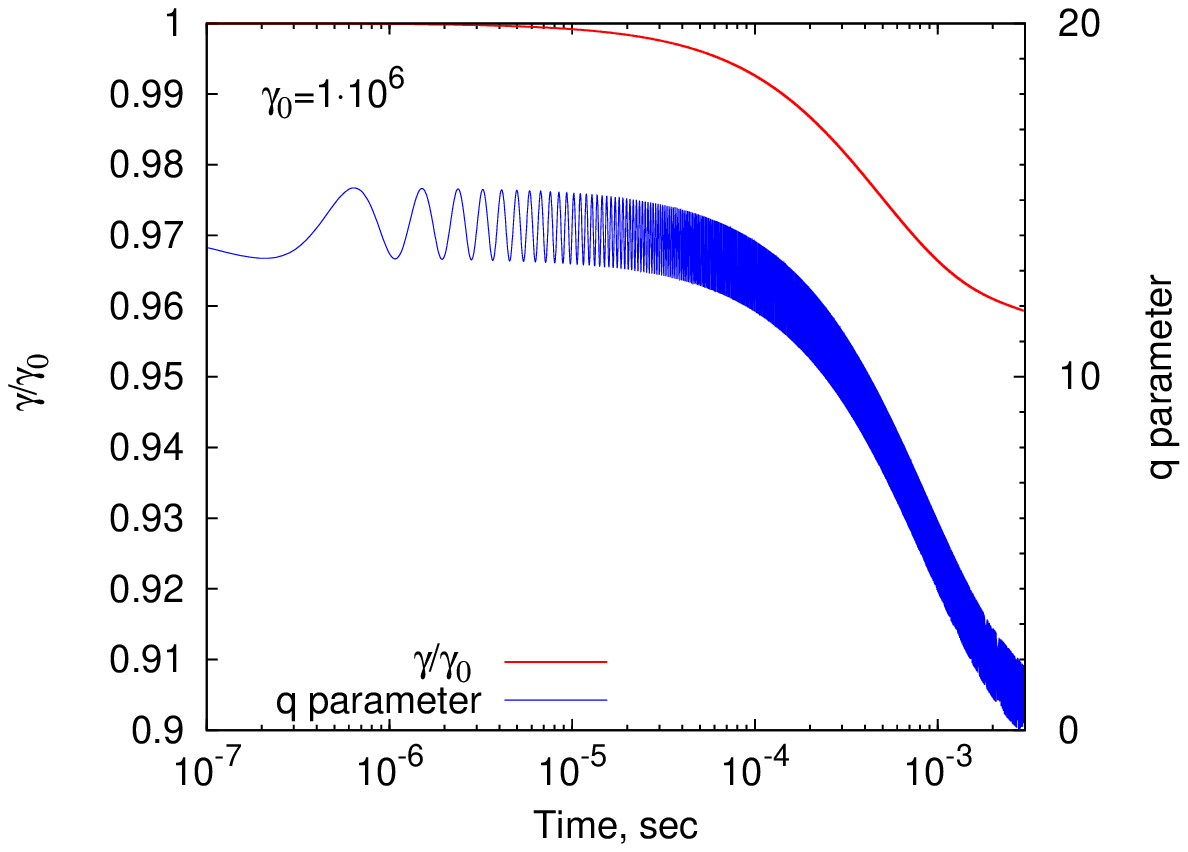} &
\includegraphics[width=0.25\textwidth]{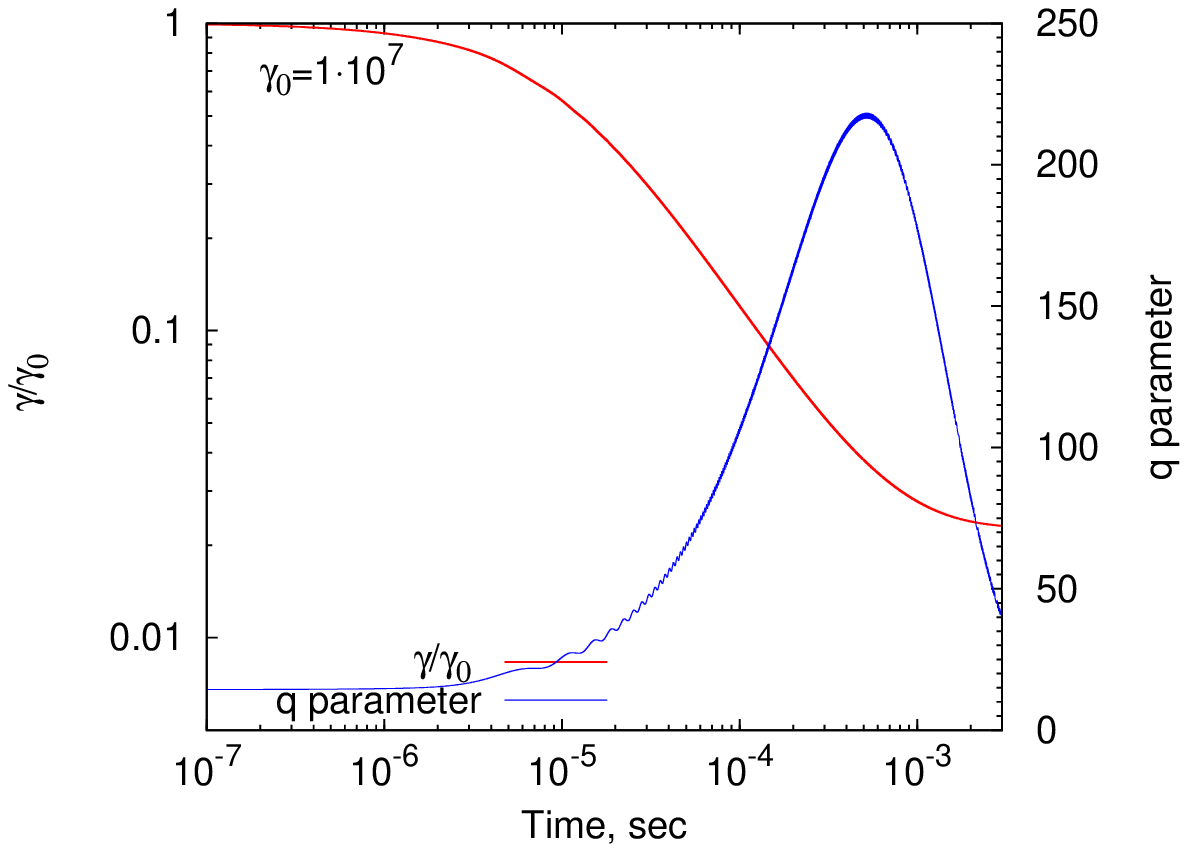} \\
\includegraphics[width=0.25\textwidth]{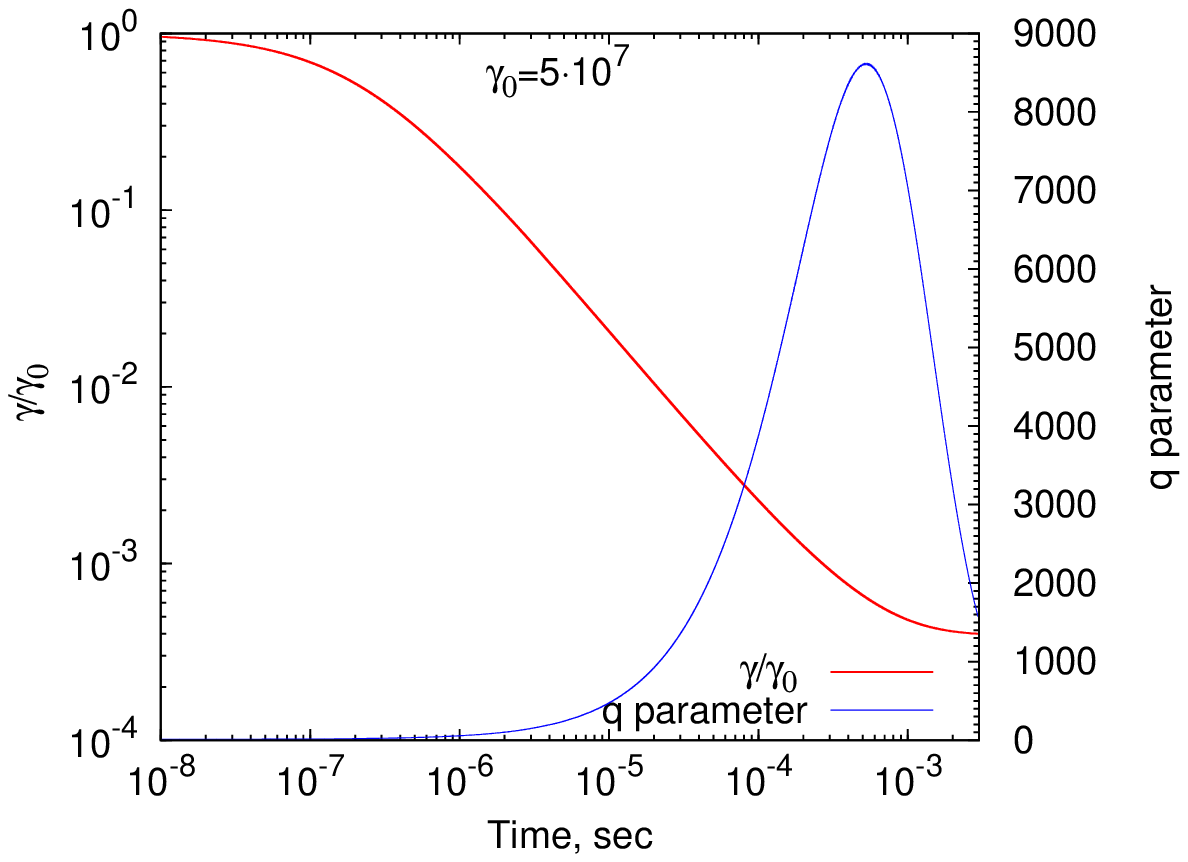} &
\includegraphics[width=0.25\textwidth]{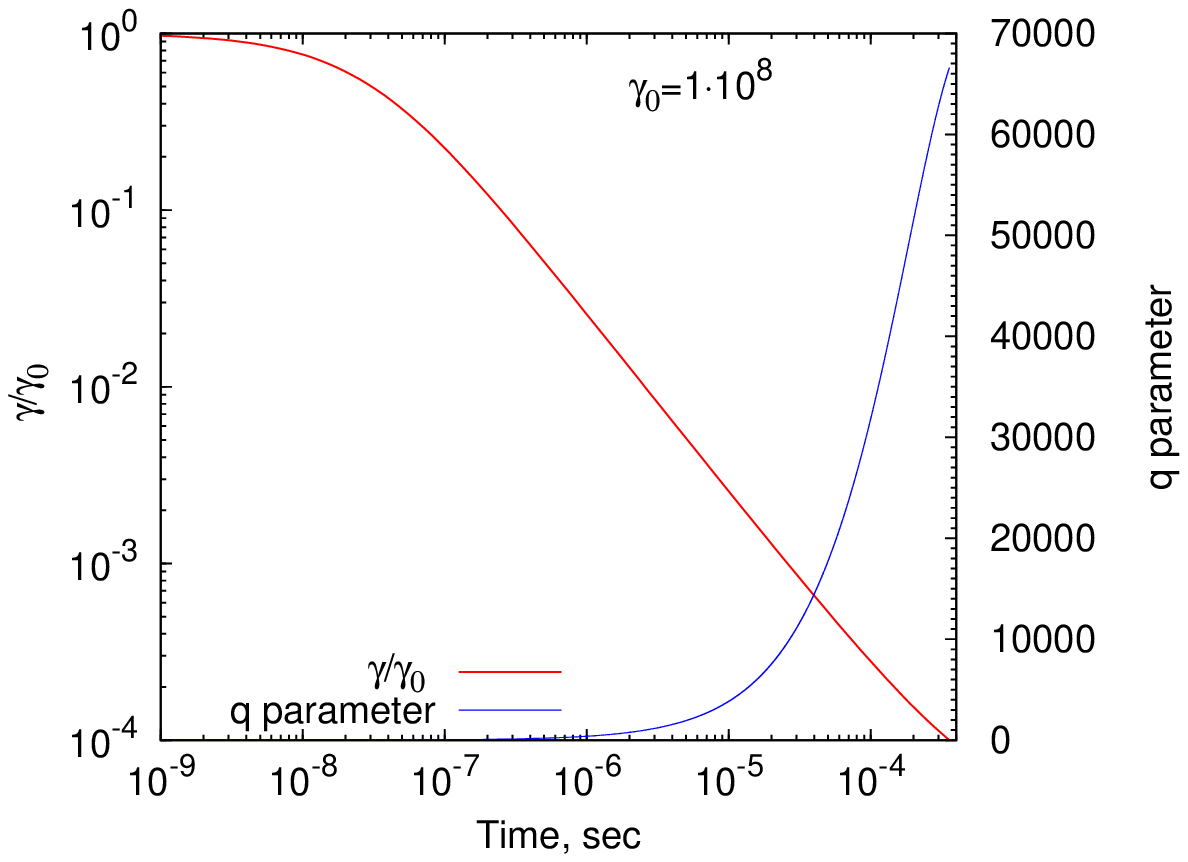}
\end{array}$
\end{center}
\caption{\label{fig:OGradComA} 
The same as in Fig.\ref{fig:OGradComD}, but for the initial direction of electrons 
at an angle $10\beta_D$.}
\end{figure}

The spectra and the radiation regimes have been studied for different initial directions and Lorenz factors of electrons.
The resulting spectra are shown in  Fig.~\ref{fig:OGrad}. The complementary plots in
Figs.~\ref{fig:OGradComD}, \ref{fig:OGradComB}, and \ref{fig:OGradComA} demonstrate the time-evolution
of the value of $q$-parameter (blue lines, right scale) and the Lorenz factor normalized  the initial Lorenz factor. 
The local behaviour of the $q$-parameter agrees well with  Eq.~(\ref{eq:qloc}) obtained from the local solution.
As discussed above, the oscillations caused by the particle gyration are within
the bandwidth of $\leq 2$. The complementary plots allow us to 
observe simultaneously the energy lose rate and the
regime of the radiation presented by the value of the $q$-parameter. If the energy loss rate is low,  the
rarefied output of the points  produces  intermittent curves for the $q$-parameter.

It is interesting  to examine   the  statement   that the particle moving along drift trajectory 
yields  minimum  intensity of 
radiation and produces less energetic photons. To do this,  the initial velocity is deflected by the angle
$\beta_D$ from the magnetic field line towards the binormal vector. The corresponding cumulative (integrated
along trajectory)  energy spectra of radiation are shown in Fig.~\ref{fig:OGrad}. The complementary plots are
shown in Fig.~\ref{fig:OGradComD}.  We can see that indeed for the initial Lorenz factors $\gamma_0=10^6$ and
$\gamma_0=10^7$ the radiation is less energetic compared to other cases  and the rates of energy losses are 
minimal as well (compare with corresponding plots in
Figs.~\ref{fig:OGradComB} and \ref{fig:OGradComA}).

The situation  however  is different for larger initial Lorentz factors;  see the curves corresponding to 
$\gamma_0=5\cdot 10^7$ and $\gamma_0=10^8$. It is seen that for the initial direction along the drift trajectory the radiation spectra extend to higher energies
than in the case of the initial direction along the magnetic field. This can be explained by the 
very intensive energy losses occurring even before the particle has made the first gyration (the first
oscillation of $q$-parameter). During this time $q\approx 1$ for the initial direction along the drift
trajectory and $q\approx 0$ for the initial direction along the magnetic field.
Eqs.~(\ref{eq:InRad})-(\ref{eq:Grad}) show that  larger values of $q$ give a more energetic radiation. 
Correspondingly, more  energetic radiation is produced in the case of the initial direction along the drift trajectory.
However,  after many gyrations the energy losses in the case of the initial direction along drift trajectory become, as expected, less intensive
compared to the case of the initial direction along magnetic field line. 

We call the attention of the reader to  the regimes of the radiation demonstrated by 
curves  in  Fig.~\ref{fig:OGradComD}.  For $\gamma_0=10^6$
and $\gamma_0=10^7$, the  $q$-parameter equals unity indicating that the radiation proceeds 
in the curvature regime. However  $q=1$  is not exact as it is demonstrated for $\gamma_0=10^7$ where the scale
for $q$-parameter slightly oscillates around unity. These small oscillations correspond to the fine gyration
around the drift trajectory. As discussed above (see Eq.~\ref{eq:applim}) the presence of such fine perpendicular motions does not influence on the applicability limits.

In the cases $\gamma_0=5\cdot 10^7$ and $\gamma_0=10^8$
the $q$-parameter has more complex behaviour. The increase at the beginning is defined mostly
by the fast energy losses. The decrease is defined by the combination of several  factors such as the reduction 
of the magnetic field strength and the change of its curvature. The increase of $q$-parameter indicates
that the radiation occurs in the synchro-curvature or the synchrotron regimes when the radiation due to
curvature of the magnetic field line is less important (for $\gamma_0=5\cdot 10^7$) or simply negligible  (for
$\gamma_0=10^8$). According to Eq.~(\ref{eq:Grad}),  the energy of the radiation maximum scales as 
$\sim\gamma^3 q$. The $q$-parameter reaches the maximum when the considerable amount of energy has been lost. 
Therefore, in spite of large $q$,  the peak of radiation shifts towards  low energies  and does not affect the cumulative
spectrum. The interesting feature can be seen at first moments when the energy oscillates with
$q$-parameter and the minimums  of $q$-parameter correspond to flatter parts of the Lorenz factor evolution curve.

The radiation spectra of electrons  launched along the magnetic field line are slightly more energetic except for 
$\gamma_0=5\cdot 10^7$ and $\gamma_0=10^8$. The reason is the same as  discussed above.
Initially, the  radiation for $\gamma_0=10^6$ and $\gamma_0=10^7$ is in transition regime, although for
$\gamma_0=10^6$ the most of the energy is lost in the curvature regime which occurs fast. Therefore the spectrum
in this case almost coincides with the spectrum for the initial direction along the drift trajectory.
For $\gamma_0=10^7$ the most of the energy is lost in the transition regime, thus the spectrum is shifted
to higher energies by $q\approx 2$.

For illustration of the effect related to initial pitch angles larger than $\beta_D$,  we show the 
case of  initial direction deflected at the angle $10\beta_D$ from the magnetic field line towards
the direction opposite to the normal vector. The corresponding spectra indicated in Fig.~\ref{fig:OGrad}
by dash-dotted lines are shifted towards higher energies. Although the $q$-parameter (Fig.~\ref{fig:OGradComA})
reaches large values, the most of the energy is lost at initial stages at $q\approx 10$. Thus the spectra
are shifted by $q\approx 10$ compared to the curvature radiation spectra.

We should note  that  the energy spectra of radiation  produced in all regimes 
contain  an exponential cut-off at highest energies  similar to the spectrum of the 
small-angle synchrotron radiation.
However,  since the radiation spectrum is very sensitive 
to the pitch-angle,  a population of electrons with
similar energies   but  different angles can result in  a superposition spectrum with a less abrupt cut-off.
The condition for  realization  of such a spectrum is that the distribution of electrons over pitch-angles
around zero angle should be wider than $\beta_D$.
   
\subsection{Polar Cap}
In the polar cap model  electrons  radiate in the region located close to the surface of  the neutron star
where the magnetic field  is much stronger  than in the outer gap model, approaching to $B\approx 10^{12}$ G.
This results in much faster damping of the perpendicular component of  motion.
The very small drift velocity $\beta_D$  implies that the drift
trajectory and the magnetic field line almost coincide, thus  even a small deflection from the magnetic field line
produces radiation quite different from the curvature radiation. However, the transition to the curvature radiation
regime occurs very fast. In the curvature regime, electrons  radiate more energetic photons than in 
the outer gap model. But the curvature of the magnetic field lines in the polar cap model with 
$\theta_0\sim 1^{\circ}$ is only
an order of magnitude larger than in the outer gap since the curvature of the dipole magnetic field scales as
$\sim\sin\theta/r$.  Correspondingly, the maximum energy of curvature radiation at the polar cap
is only an order of magnitude higher  than in the outer gap.

\begin{figure}
 \begin{center}
  \includegraphics[width=0.5\textwidth]{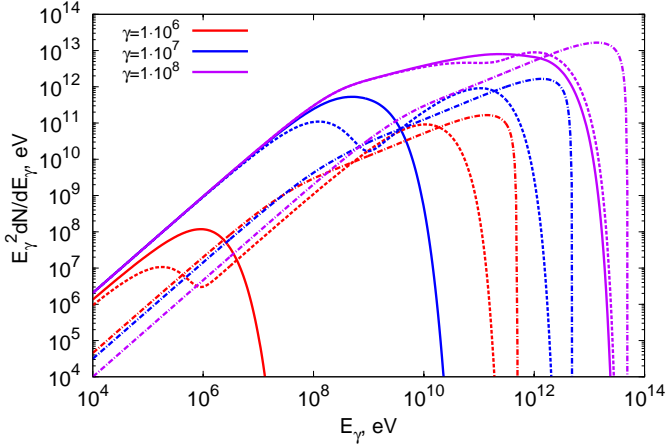}
  \caption{\label{fig:PCrad}
 The cumulative (integrated
  along trajectory) radiation spectra of electrons calculated for the 
  outer gap model of the pulsar magnetosphere. The curves are obtained  
  for different initial Lorenz factors  of electrons  $\gamma=10^6, 10^7, 10^8$, and
  for different initial directions relative to the magnetic field lines:   
  along the magnetic field line (dashed lines), and for two pitch angles $1/\gamma_0$ (dashed lines) 
  and $100/\gamma_0$  (dashed-dotted lines).}
 \end{center}
\end{figure}

\begin{figure}
\begin{center}$
\begin{array}{cc}
\includegraphics[width=0.25\textwidth]{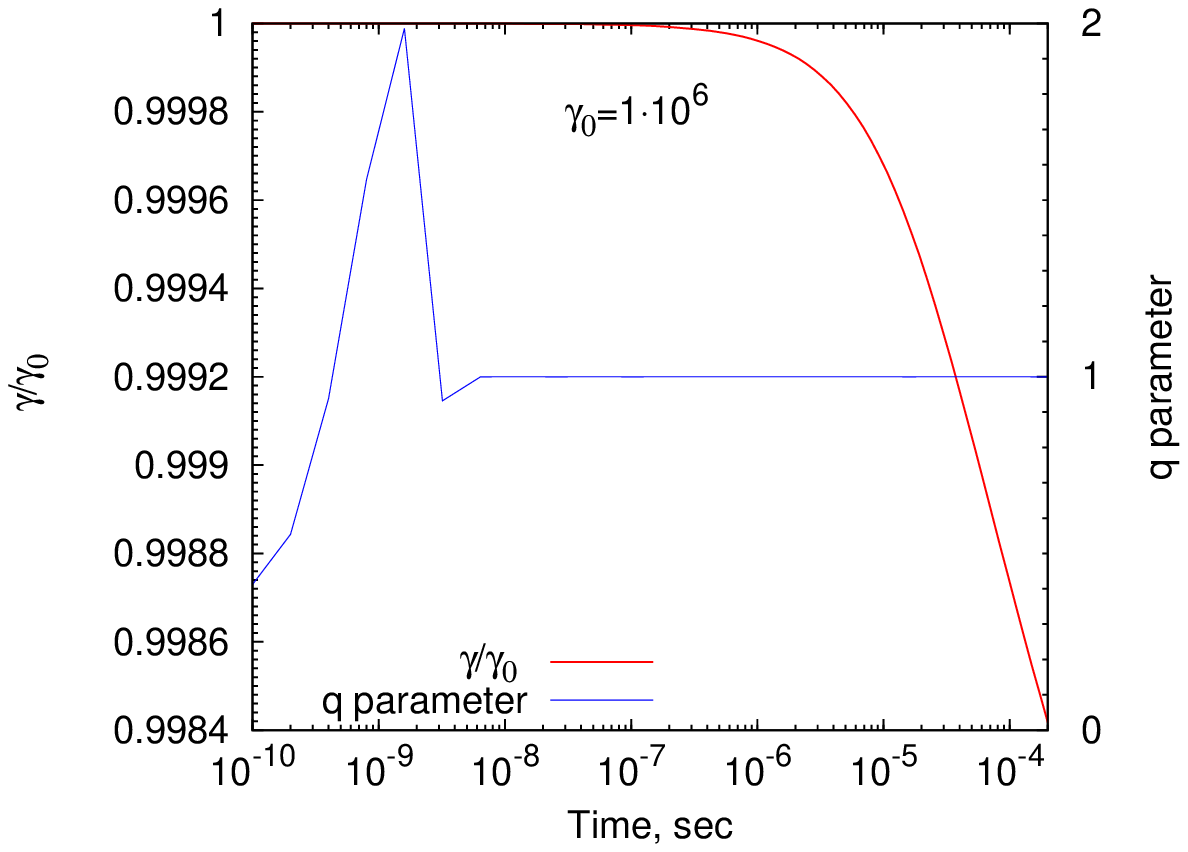} &
\includegraphics[width=0.25\textwidth]{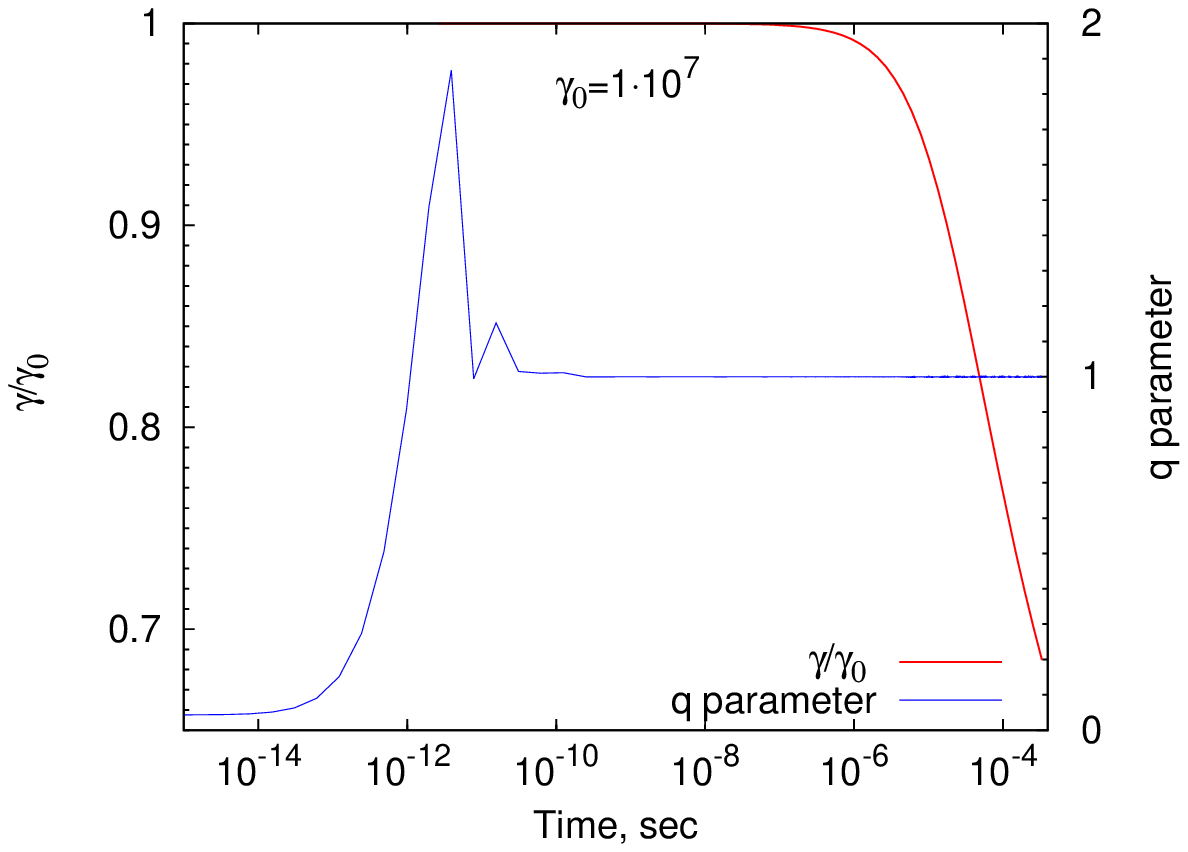}
\end{array}$
\includegraphics[width=0.25\textwidth]{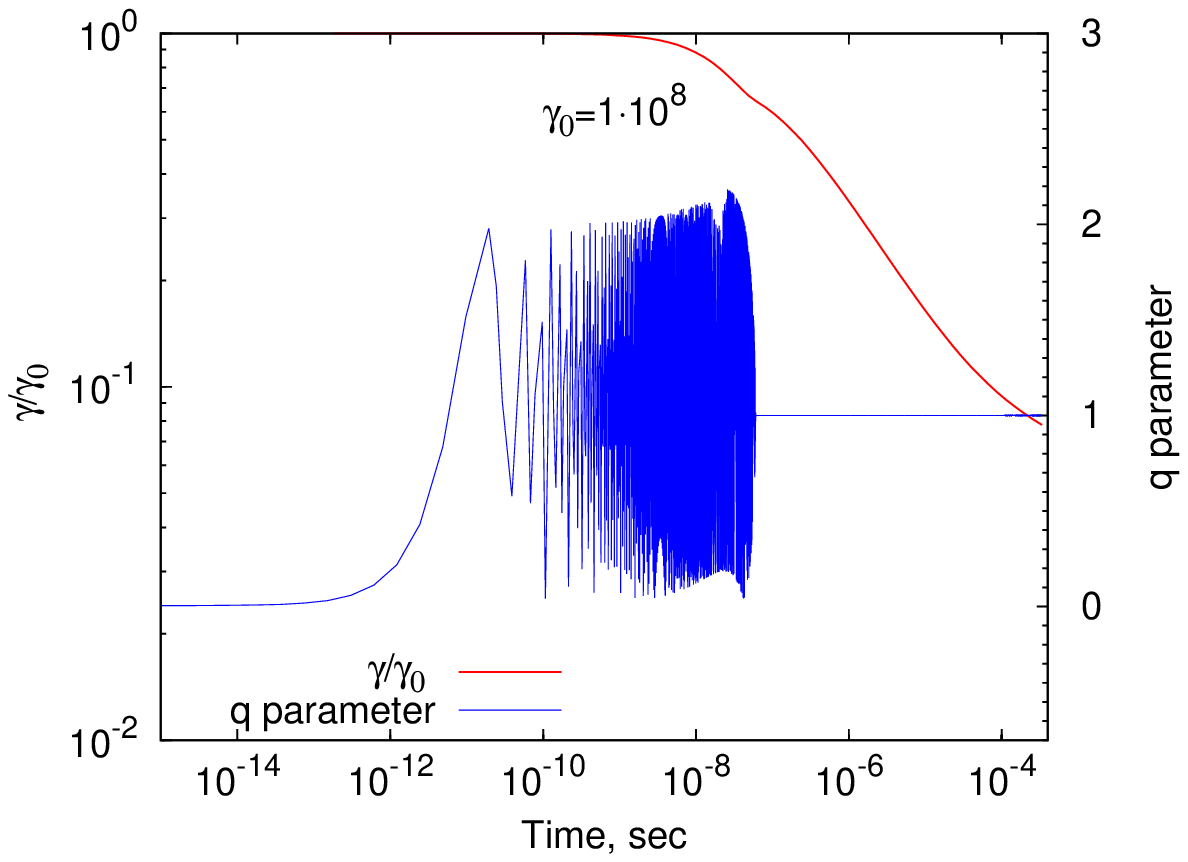}
\end{center}
\caption{\label{fig:PCalong}
Time evolution of the   $q$-parameter and the electron
 Lorenz factor in the outer gap model (complementary to Fig.~\ref{fig:PCrad}). Three  panels correspond to 
 the initial Lorenz factor of  electrons  $\gamma=10^6, 10^7,  10^8$
 and their initial direction  along the  magnetic field line.}
\end{figure}

\begin{figure}
\begin{center}$
\begin{array}{cc}
\includegraphics[width=0.25\textwidth]{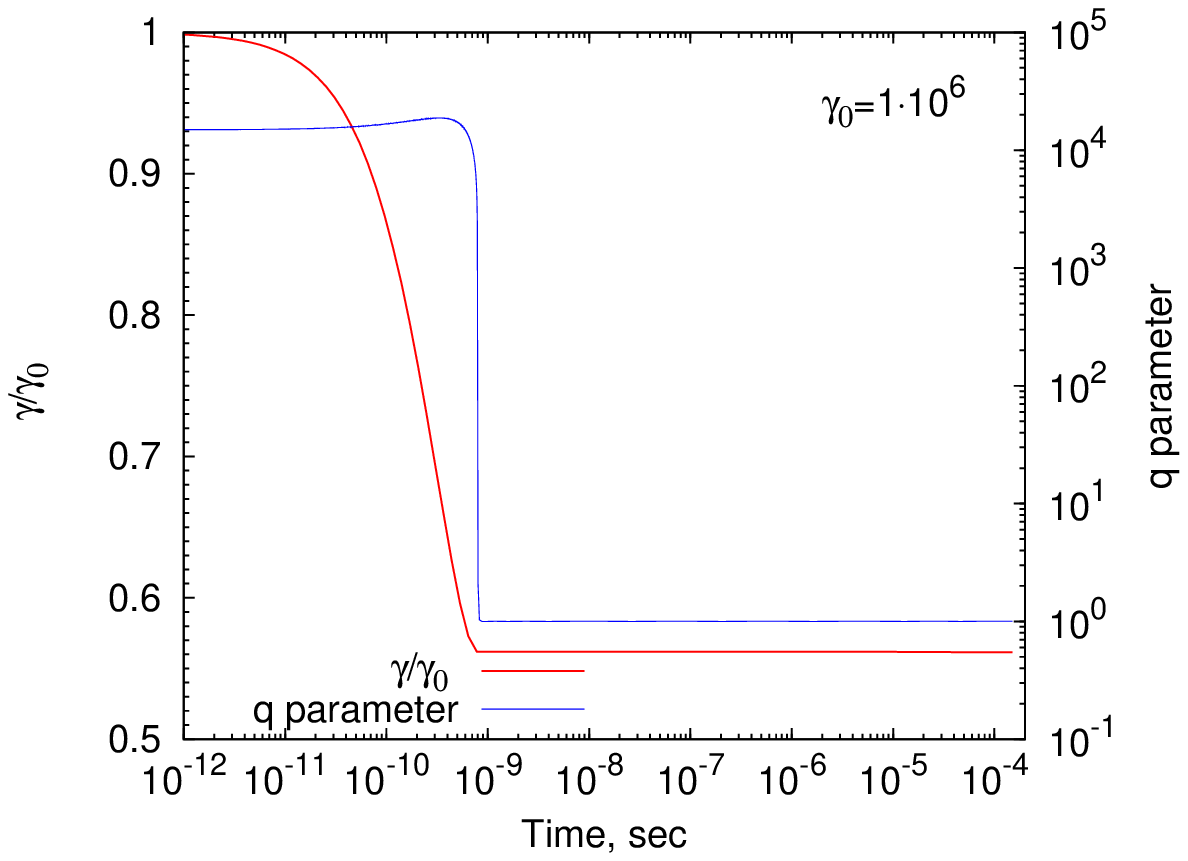} &
\includegraphics[width=0.25\textwidth]{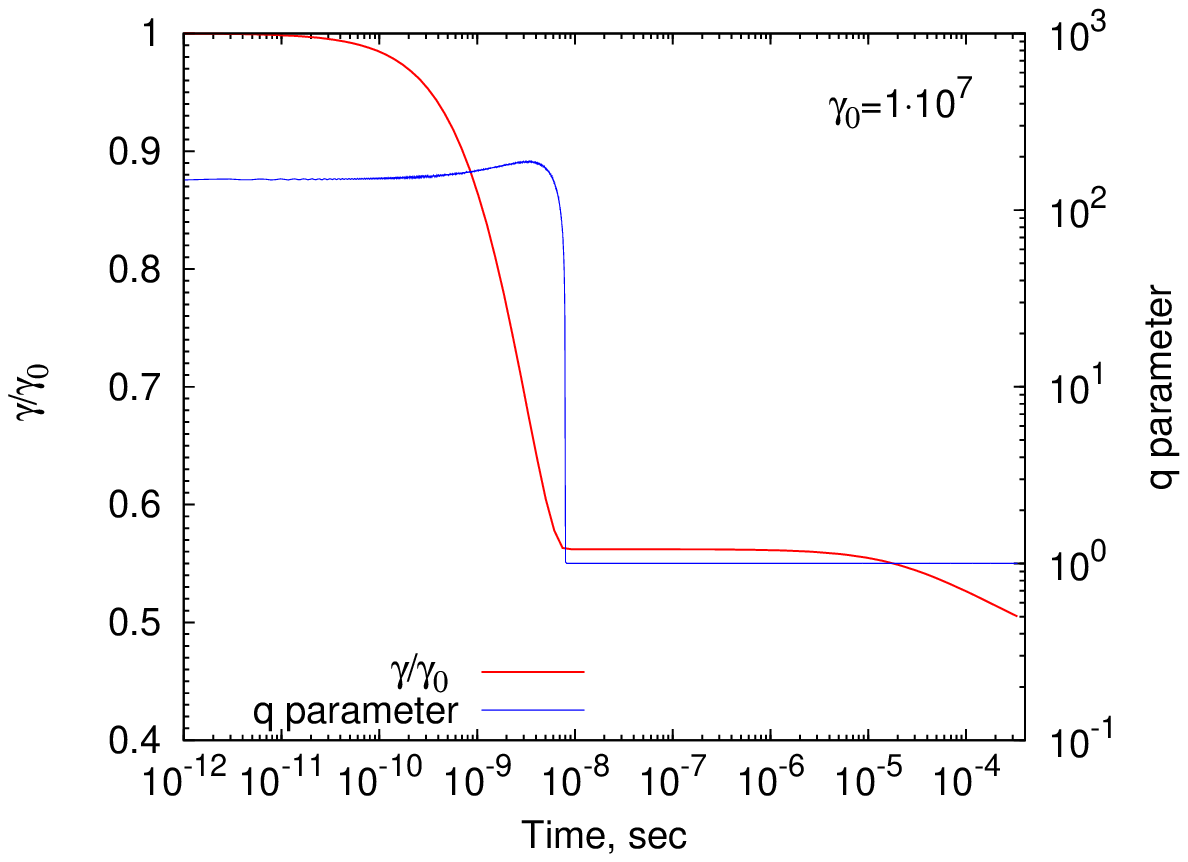}
\end{array}$
\includegraphics[width=0.25\textwidth]{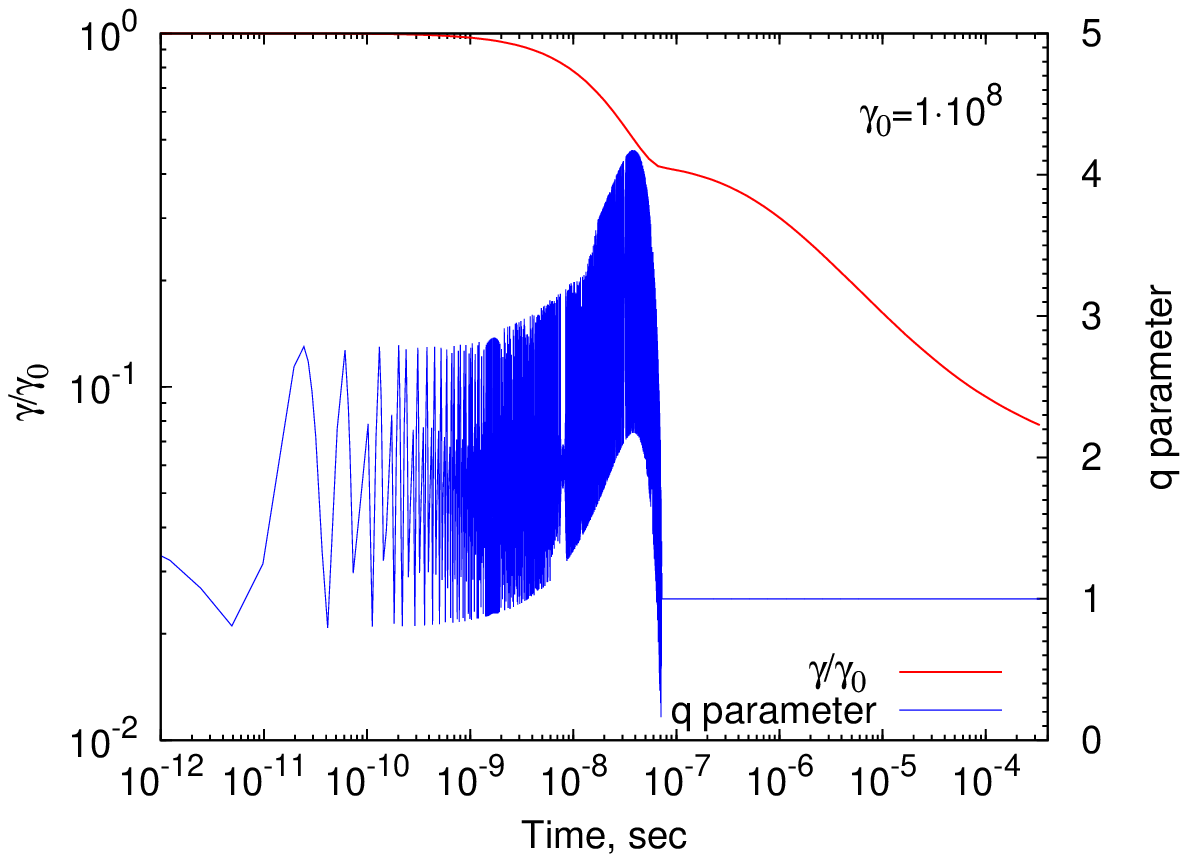}
\end{center}
\caption{\label{fig:PC1g}
The same as in Fig.\ref{fig:PCalong}, but for the initial direction of electrons 
at pitch angle  $1/\gamma_0$.}
\end{figure}

\begin{figure}
\begin{center}$
\begin{array}{cc}
\includegraphics[width=0.25\textwidth]{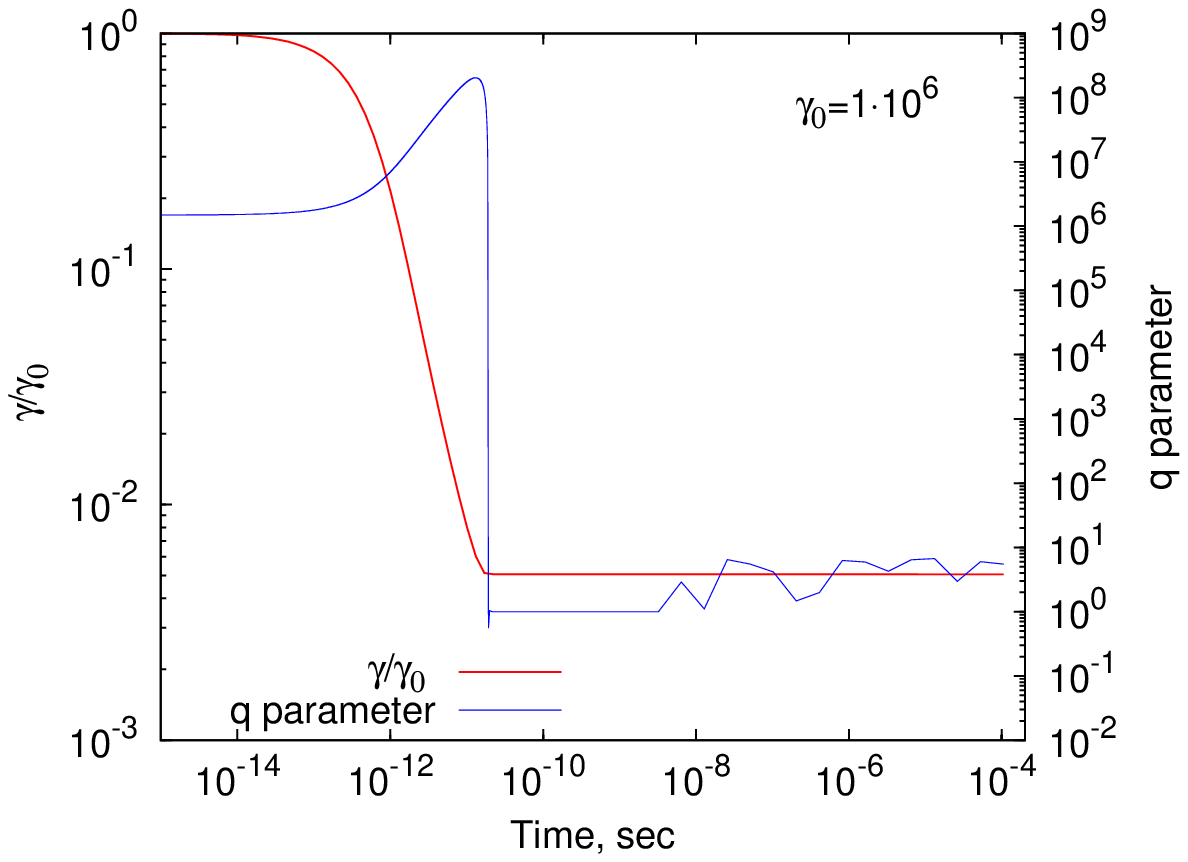} &
\includegraphics[width=0.25\textwidth]{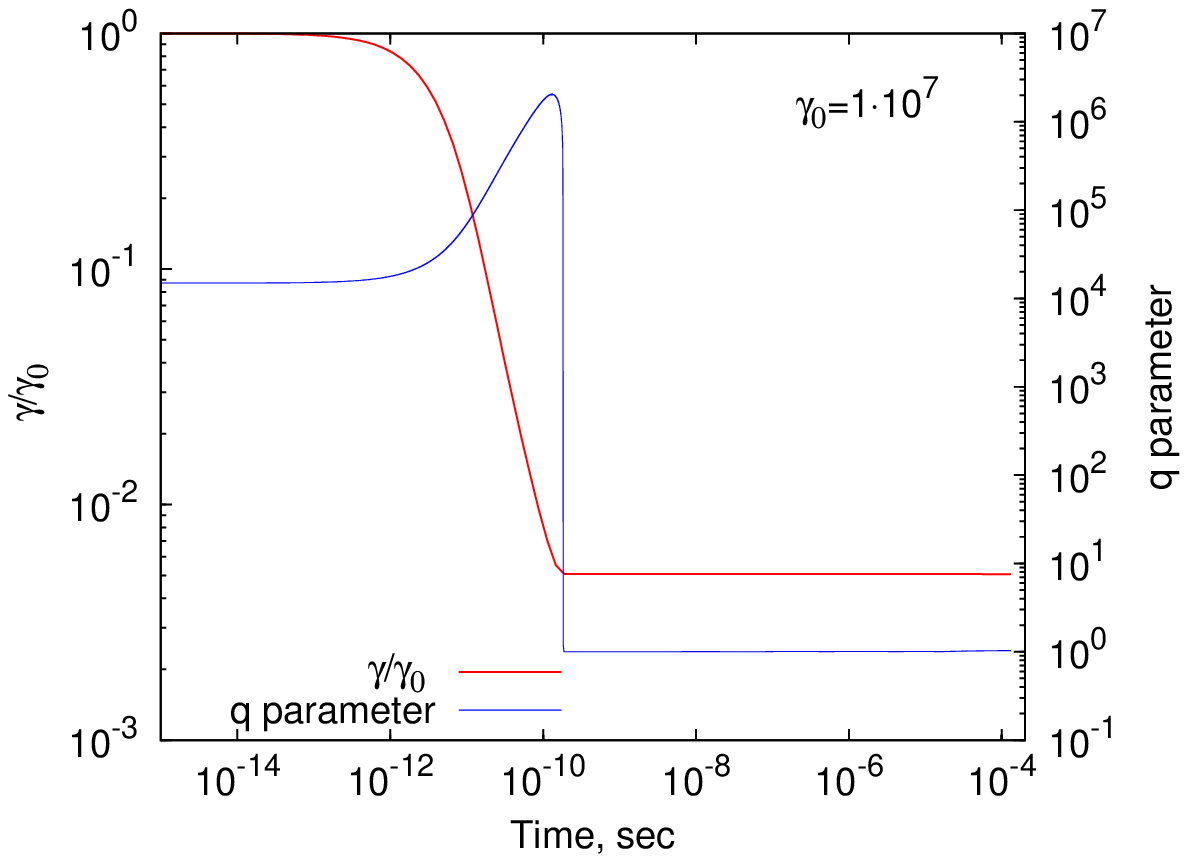} 
\end{array}$
\includegraphics[width=0.25\textwidth]{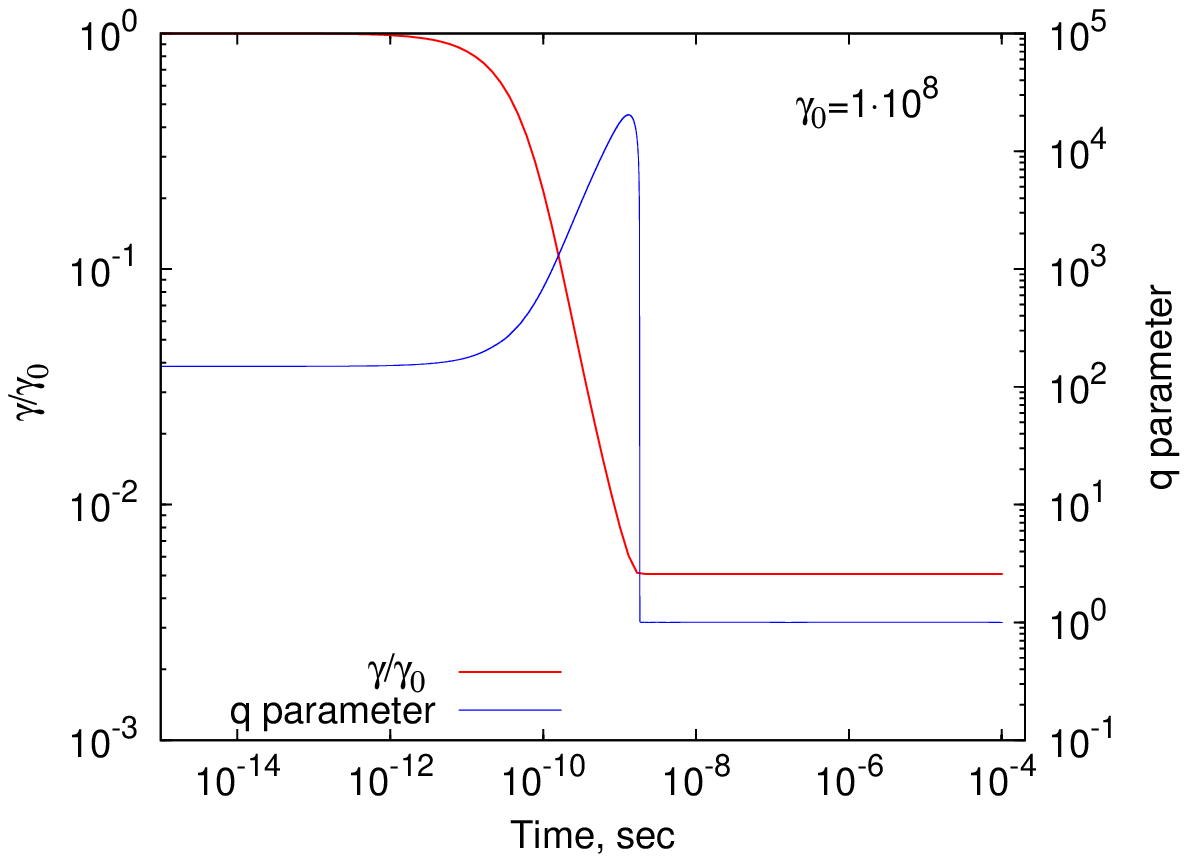}
\end{center}
\caption{\label{fig:PC100g}
The same as in Fig.\ref{fig:PCalong}, but for the initial direction of electrons 
at pitch angle  $100/\gamma_0$.}
\end{figure}

The  energy spectra of radiation calculated  for the polar cap model are shown  in Fig.~\ref{fig:PCrad}. The complementary plots for the
evolution of the $q$-parameter and the electron Lorenz factor are presented in Figs.~\ref{fig:PCalong}, \ref{fig:PC1g}, and
\ref{fig:PC100g}. The initial position of the particle is $R_0=10^6$ cm,  and $\theta_0=3^{\circ}$.
The spectra indicated by solid lines correspond to the case when the initial direction of 
particle is along magnetic field line. In this case the radiation is  in the curvature regime.  
But at the initial stage,  the radiation  proceeds in a very fast synchro-curvature regime, but
abruptly turns to the regime  with $q \approx 1$ and fast   
oscillations caused by fine gyrations.

The abrupt change of the regimes leads to an  interesting feature in  the cumulative 
spectra for the case of an initial pitch
angle $\alpha_0=1/\gamma_0$ (dashed lines). Because of small changes in energy and fast change of the 
$q$-parameter,  the energy spectra consist of two peaks. The peak at higher energies 
 is produced by synchrotron
radiation ($q\gg 1$, see Fig.~\ref{fig:PC1g}), while the lower energy peak corresponds to curvature radiation regime.
The double-peak structure  disappears  for large  initial pitch angles. For example, for 
the pitch angle $100/\gamma_0$   the transition to the curvature regime is very fast, and  the electron 
enters into this regime with dramatically reduced Lorentz factor.  
Thus the peak of the curvature radiation not only is shifted to  smaller 
energies, but also is too weak to be seen in the  cumulative spectrum\footnote{Note that the pitch angle $100/\gamma_0$ which we treat  as `large',  still is extremely  small,  $\sim 2 (\gamma_0/10^7)^{-1}$arcsec.}.

In very  strong magnetic fields,  namely 
when the parameter $\chi=B\gamma \sin\alpha/B_{cr} \geq 1$,
the radiation is produced in the  quantum regime. 
Let's  assume that  the initial pitch angle is inverse proportional to the initial Lorentz factor, 
$\alpha_0=a/\gamma_0$. This makes the condition of radiation in the quantum regime independent on $\gamma_0$:  
\begin{equation}
\chi=\frac{B}{B_{cr}}\gamma \sin{\alpha_0}=\frac{1\cdot 10^{12} G}{2.94\cdot 10^{13} G} a \approx 3.4\cdot 10^{-2} a.
\end{equation}
Thus,  at the initial pitch angle with $a>30$,  the electrons radiate in quantum regime.
 Dash-dotted lines in Fig.~\ref{fig:PCrad} present radiation spectra for the initial pitch-angles
$\alpha_0=100/\gamma_0$. Note that in the quantum regime almost the entire energy of the parent electron is transferred to the radiated photon. Thus we should expect abrupt cutoff in the radiation spectra. This effect is clearly  seen in   Fig.~\ref{fig:PCrad} (dot-dashed curves corresponding to the initial pitch angle 
$\alpha_0=100/\gamma_0$). The gamma rays produced in the quantum regime are sufficiently energetic to be 
absorbed in the magnetic field through the  $e^+e^-$  pair production.  This will lead to the development of an electromagnetic cascade in the magnetic field. The spectrum of cascade gamma-rays that escape  the pulsar magnetosphere will be quite different from the spectra shown in Fig.~\ref{fig:PCrad}. 

\subsection{Protons in the Black Hole magnetosphere}
The acceleration of the protons to ultrahigh energies in the potential gap of the spinning supermassive black hole 
should be accompanied by curvature radiation in the magnetic field which threads the
black hole (\cite{Levinson2000}, \cite{Aharonian2002}). 
Here we briefly  examine the radiation regimes and the gamma ray spectra of accelerated protons produced in this model.  As before,  we consider the radiation in the dipole magnetic field adopting the following (typical for a SMBH)
parameters $B_{*}=10^4$ G and $R_{*}=10^{14}$ cm. The initial position
of the particle is set at the polar angle $\theta_0=5^\circ$ relative to the magnetic dipole axis. 

The larger (compared to electron)  mass of proton  leads to
the larger drift velocity $\beta_D$.
Therefore the spectra and the radiation regimes in the 
case of protons are  less sensitive to the initial direction of motion. 
The radiation of protons deviates from the curvature radiation when the perpendicular component of 
proton's  Lorentz factor exceed     
\begin{equation}
\gamma_{\perp}=\gamma\beta_{D}=\frac{m_{p}c^2\gamma^2}{eBr_0}=2.7\cdot 10^{7}\left(\frac{\gamma}{10^{10}}\right)^2 \ .
\end{equation}
This is larger by the factor of $m_p/m_e\approx 2\cdot 10^3$  compared to the  same  condition for electrons. 
It is interesting to note that for  the same Lorenz factor,  in the synchrotron regime  protons radiates 
much weaker than electrons,  whereas in the curvature regime they radiate equally. 

\begin{figure}
 \begin{center}
  \includegraphics[width=0.5\textwidth]{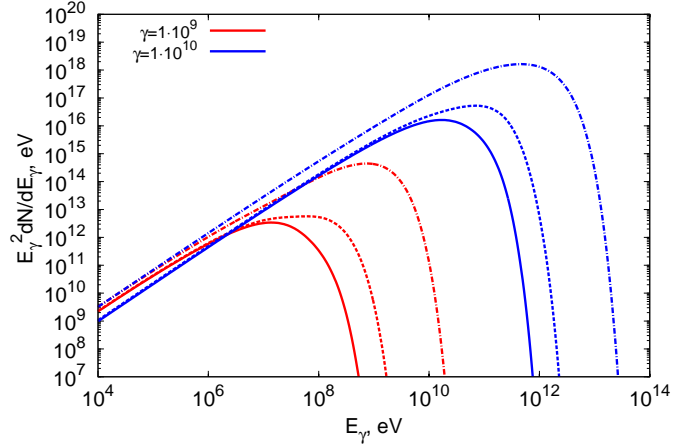}
  \caption{\label{fig:BHrad} The cumulative (integrated
  along trajectory) radiation spectra of protons  in the magnetic field of rotating supermassive black hole. The chosen parameters are described in the text.   The spectra correspond to two different Lorentz factors and  three different initial pitch angles 
      relative to the magnetic field lines:  $10^{-4}$ (solid), $10^{-3}$ (dashed), $10^{-2}$ (dash-dotted) radian for  $\gamma_0=10^{9}$ (red) and  $10^{-3}$ (solid), $10^{-2}$ (dashed), $10^{-1}$ (dash-dotted) radian for $\gamma_0=10^{10}$ (blue).}
 \end{center}
\end{figure}

\begin{figure}
\begin{center}$
\begin{array}{cc}
\includegraphics[width=0.25\textwidth]{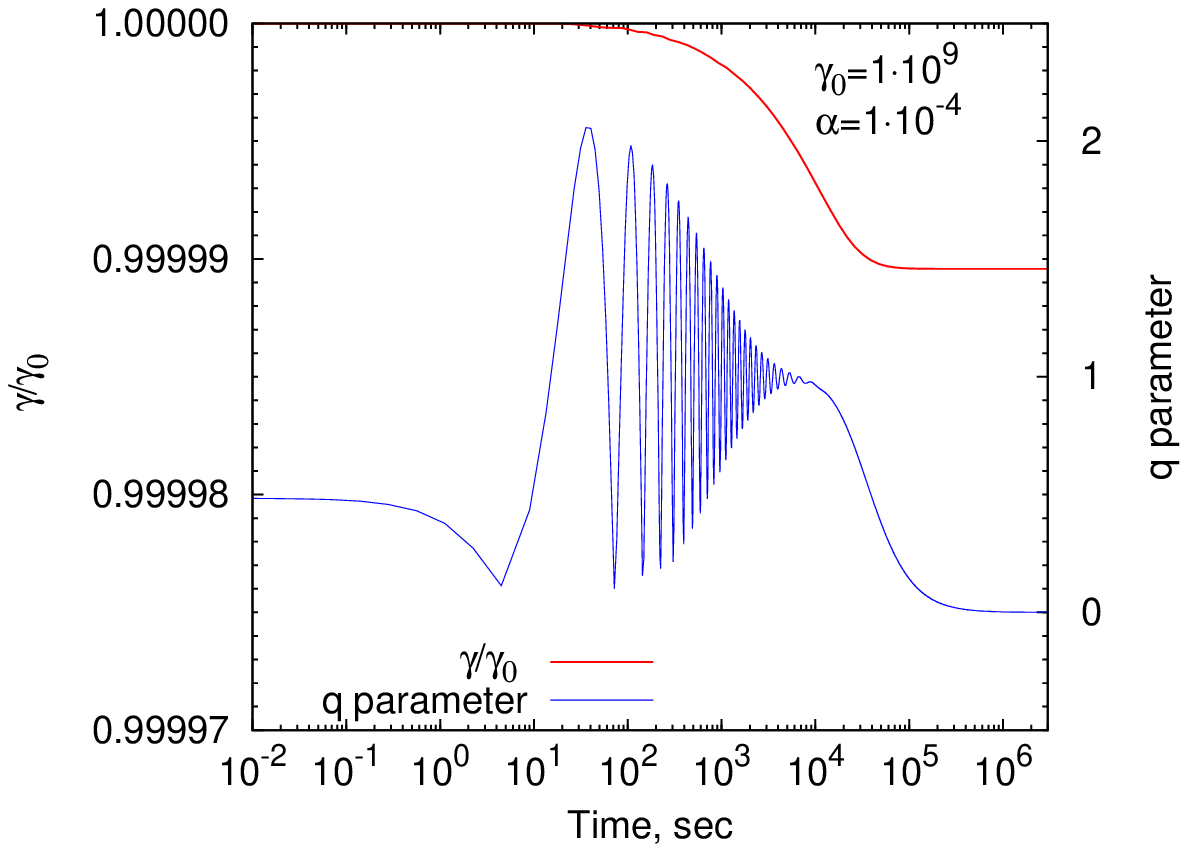} &
\includegraphics[width=0.25\textwidth]{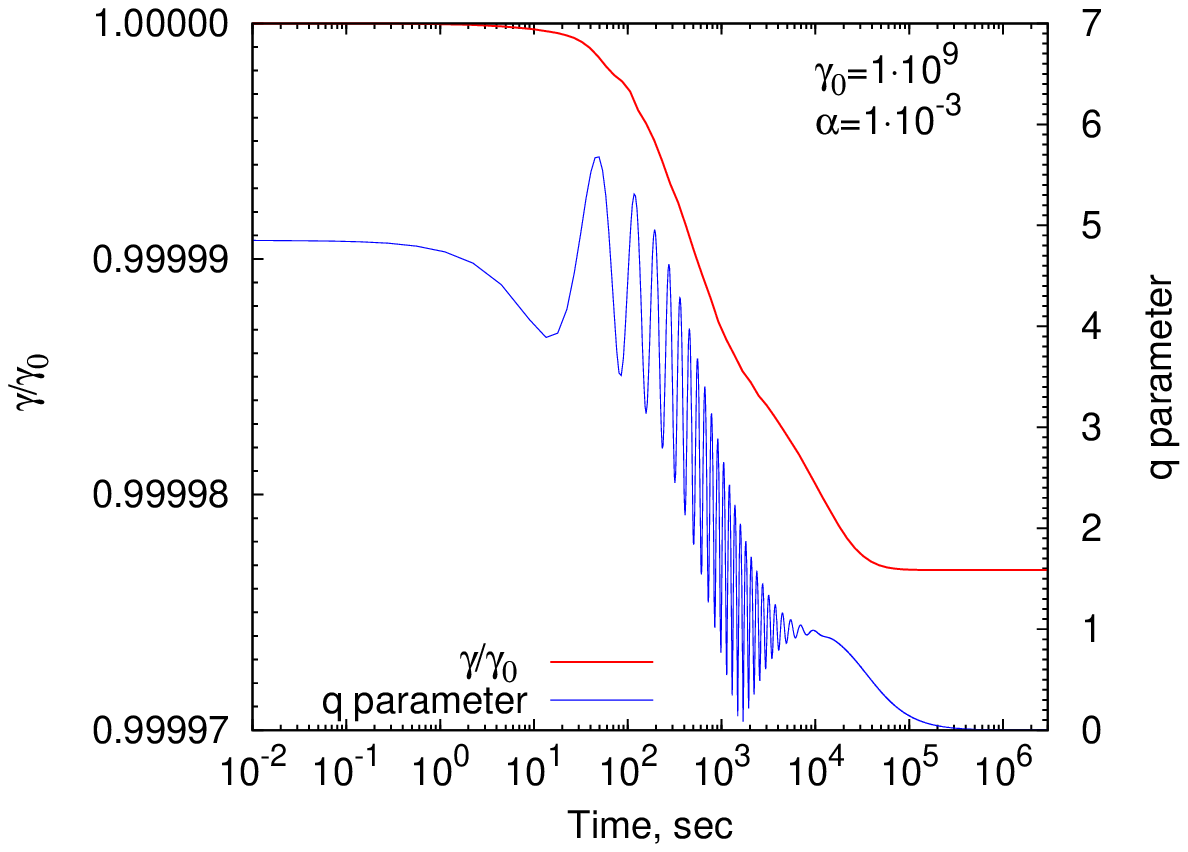}
\end{array}$
\includegraphics[width=0.25\textwidth]{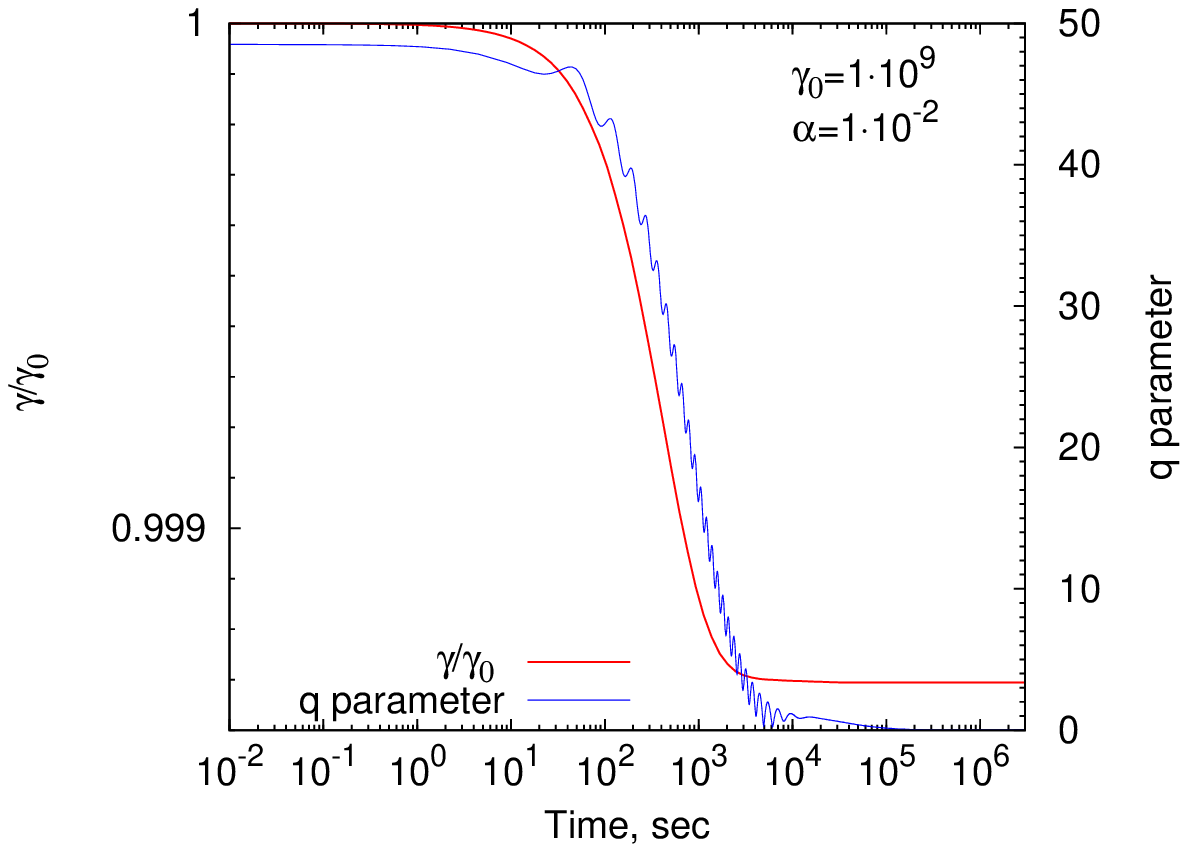}
\end{center}
\caption{\label{fig:BHradg9} Time-evolution of the Lorenz factor and the $q$-parameter
in the magnetic field of a supermassive black hole with the 
initial Lorenz factor of protons $\gamma_{0}=10^9$ and their 
initial pitch angles $\alpha=10^{-4}$, $10^{-3}$, and $10^{-2}$.}
\end{figure}

\begin{figure}
\begin{center}$
\begin{array}{cc}
\includegraphics[width=0.25\textwidth]{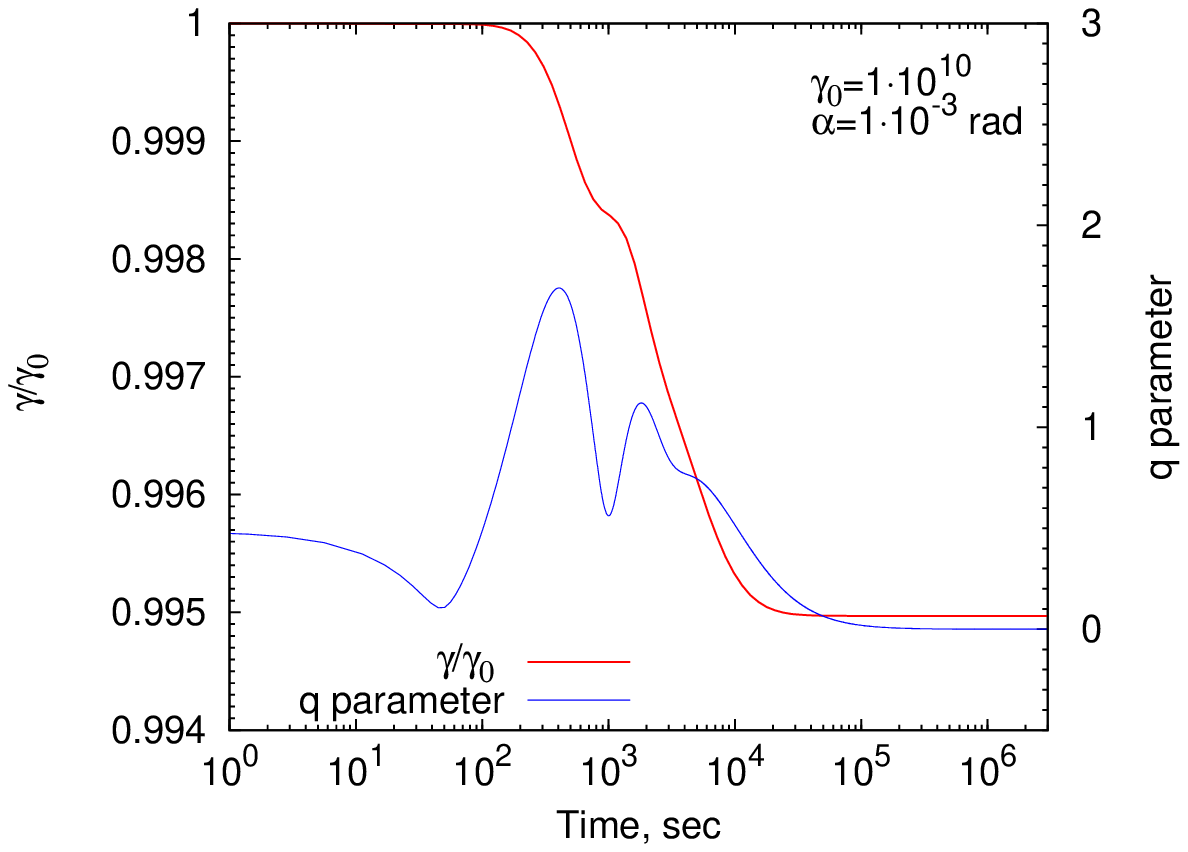} &
\includegraphics[width=0.25\textwidth]{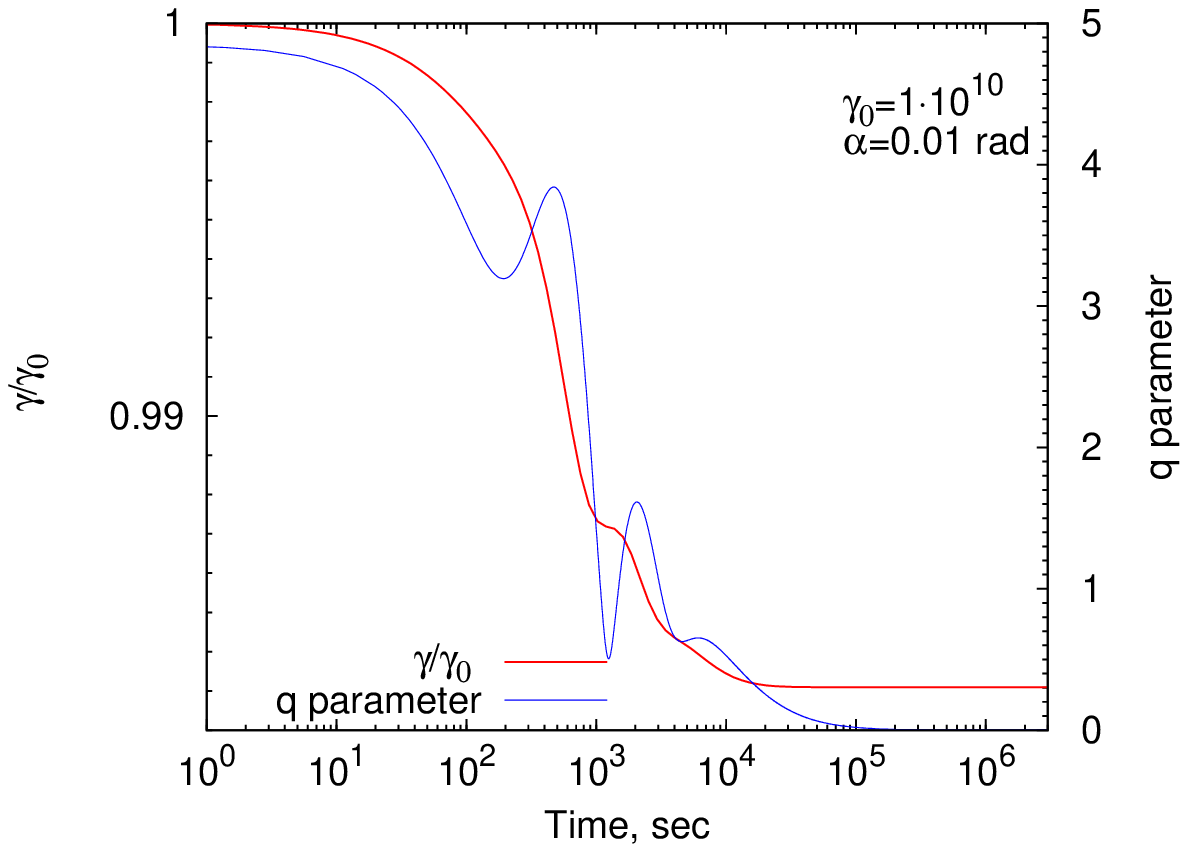}
\end{array}$
\includegraphics[width=0.25\textwidth]{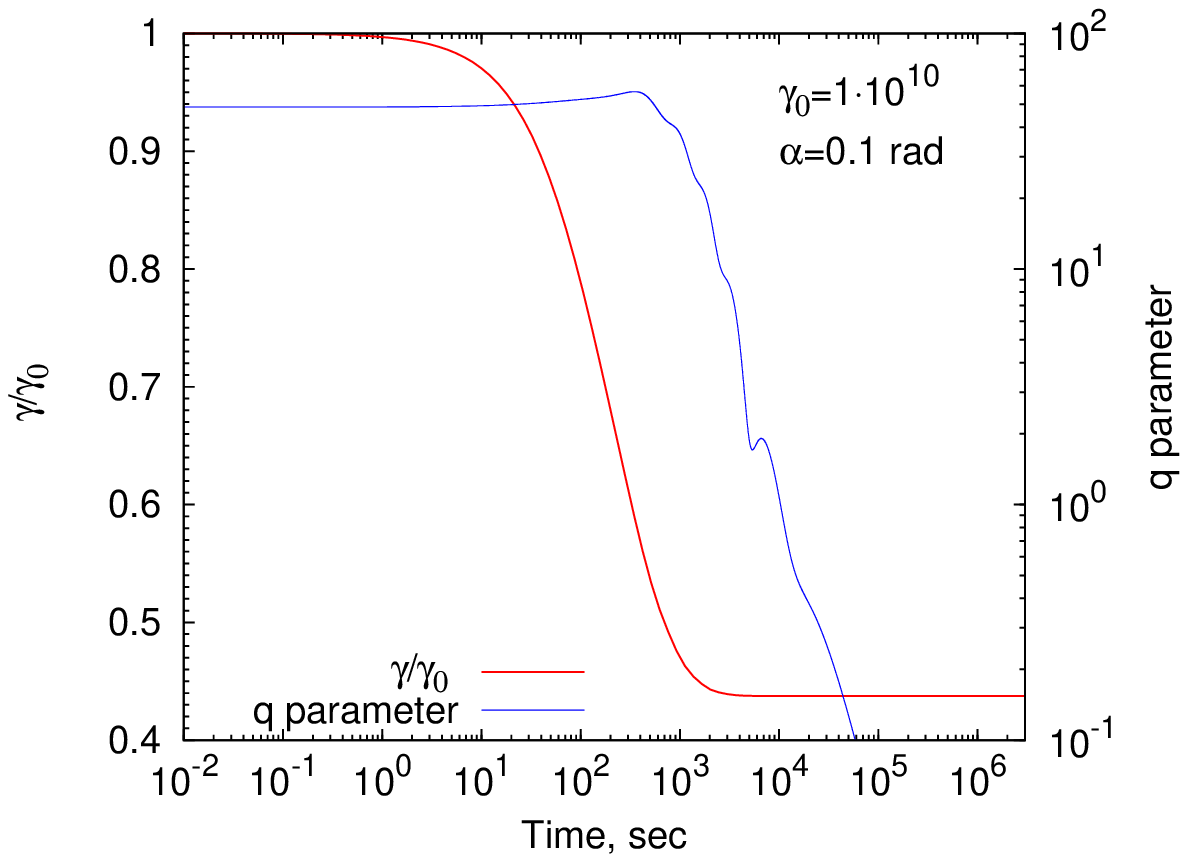}
\end{center}
\caption{\label{fig:BHradg10} Time-evolution of the Lorenz factor and the $q$-parameter
in the magnetic field of a supermassive black hole with the 
initial Lorenz factor of protons $\gamma_{0}=10^{10}$ and their 
initial pitch angles $\alpha=10^{-3}$, $10^{-3}$, and $10^{-1}$.}
\end{figure}

The radiation spectra of protons with different initial pitch
angles are presented in Fig.~{\ref{fig:BHrad}}. The smallest angle $\alpha_0$ is close to  $\beta_D$,  and the
proton radiates in the synchro-curvature regime (solid lines). Because of high energy the gyration period
is large, and in the case of $\gamma_0=10^{10}$ the particle makes only several gyrations before escaping 
the region with high magnetic field (see Fig.~\ref{fig:BHradg10}). After that,  the proton radiates with a very low rate, 
and the $q$-parameter  approaches zero. In the case of $\gamma_0=10^{9}$,  the proton  gyrates more frequently  (see
Fig.~\ref{fig:BHradg9}) and there is seen more pronounced  tendency of $q$  approaching to zero. The other curves  
correspond to the initial pitch angles $10\alpha_0$ and $100\alpha_0$.
The spectra are shifted towards higher energies by the same factors of $q\approx 10$ and $q\approx 100$. Accordingly, for the Lorenz factor $\gamma_0=10^{11}$ which is not shown in  
figures,  the proton will radiate predominantly in the curvature regime and the spectrum will be shifted by a factor of 
$(10^{11}/10^{10})^3=1000$ relative to the curvature spectrum of the proton with initial Lorenz factor 
$\gamma_0=10^{10}$.  Despite the small energy losses relative to the initial energy,  the
amount of the radiated energy is quite large. As in the case of the pulsar polar cap model, 
for the chosen parameters  the curvature radius of the magnetic field lines is larger by an order of magnitude compared to  the gravitational radius of the black hole which usually is taken for evaluation of curvature radius. It yields  smaller energy losses and increase the maximum Lorenz factor of acceleration compared to  the case when the gravitational radius of the black hole is used as a curvature radius. 
To be more specific, the maximum Lorenz factor of a particle in radiative-loss limited regime scales with curvature radius $R_c$ as $\sqrt{R_c/R_{*}}$ (\cite{Aharonian2002}, \cite{Levinson2000}).
 
\section{Summary}

The radiation of relativistic particles in a strong curved magnetic field is of a great astrophysical interest, 
in particular in the context of magnetospheric gamma-ray emission of rotation-powered pulsars.
In these objects with  very strong magnetic fields,  the radiation proceeds in quite complex regimes, so it cannot be reduced  merely to the consideration of the  nominal synchrotron and curvature channels.  For proper understanding of  these radiation regimes and the transitions between them,  the accurate treatment  of the particle trajectory is  a key issue. It can be done by solving the equations of motion taking into account energy losses of charged particles. 
In this work we followed this approach to explore the radiation features of ultrarelativistic particles in a  dipole field which is a good approximation for the magnetic field structure in compact astrophysical objects. The accurate  numerical solutions for the test particle trajectory  allowed us to trace the radiation regimes and  calculate self-consistently the radiation spectrum without any {\it a priori} assumption concerning the radiation regime. 

In this paper we demonstrate that even  small deflections of the initial  particle motion from the magnetic field
lines may result in a radiation spectrum quite different from the spectrum the curvature radiation.  For any 
initial pitch angle,  the particle tends, while losing  its energy, to turn to the curvature radiation regime and to
move strictly along drift trajectory (although never can achieve the latter). 
In principle, the fast increase of curvature of the magnetic
field lines can turn the regimes in inverse order. 

In different environments (or different positions relative to magnetic dipole) the transition to the curvature radiation
regime proceeds  with different rate. While in the polar cap model the transition
occurs almost instantly, in the outer gap the transition could last so long that the particle
would not turn to the curvature regime while passing the gap. The spectrum of the radiation becomes very
different from the spectrum of the curvature radiation if the pitch angle of the particle is greater than  drift speed
$\beta_D=v_D/c$,  and quite similar if it is smaller than $\beta_D$. For the typical parameters of the
polar cap model,  $\beta_D=2.2\cdot 10^{-9}(\gamma/10^{8})$. This implies that  the even tiny deflection from
the magnetic field line leads to the spectrum different from the curvature radiation spectrum.  

Significant  deviations of the radiation spectra from the nominal curvature radiation spectrum is expected also in the outer gap model.   The particles do not move along magnetic field lines but gyrate around drift trajectory. This results in a  different (more ``energetic'') spectrum from the conventional curvature radiation spectrum. In the outer gap model, 
electrons  with initial direction along magnetic field line start  to radiate in the synchro-curvature regime, 
when the both the curvature of the  magnetic field and its strength play equally important role. The effect is quite significant at large Lorentz factors of electrons, $\gamma \geq 10^7$. 

Finally, we demonstrate the strong impact of the initial pitch angle on the radiation spectrum in the scenario of acceleration and motion of ultrahigh energy protons in the magnetosphere of a supermassive black hole.  
\acknowledgements{
A.P. is a fellow of the International Max Planck
Research School for Astronomy and Cosmic Physics
at the University of Heidelberg (IMPRS-HD).} 

\appendix
\section{Energy losses in the quantum regime}
In a very  strong magnetic field,  the  ultrarelativistic electrons  can radiate in the quantum regime, 
provided that
\begin{equation}
\chi=\frac{B}{B_{cr}}\gamma\sin\alpha \gtrsim 1,
\end{equation}
where $B_{cr}=\frac{2m^2c^3}{3e\hbar}\approx 2.94\cdot 10^{13}$ G. The energy lose rate 
can be written in the form \citep{Bayer,Landau4}
\begin{equation}\label{eq:QEnLoss}
\left| \frac{dE}{dt} \right|=\frac{e^2 m^2 c^3}{\sqrt{3}\pi\hbar^2} \overline{H}(\chi) \, ,  
\end{equation}
where 
\begin{equation}
\overline{H}(\chi)=\int_{0}^1 H(\tau,\chi) d\tau \ ,
\end{equation}
and  
\begin{eqnarray}
H(\tau,\chi)=\chi\left[(1-\tau)F(x)+x\tau^2K_{2/3}(x) \right],
\quad x=\frac{\tau}{\chi(1-\tau)},
\end{eqnarray}
where $K_{2/3}(x)$ is the modified Bessel function of the order $2/3$, $F(x)$ is
the emissivity function of the synchrotron radiation (see Eq.~\ref{eq:fx}), $\tau=\epsilon/E$, where
$\epsilon$ is the energy of the radiated photon, $E$ is the energy of the radiating particle.

For  calculations it is  convenient to express  $\overline{H}(\chi)$ in Eq.(\ref{eq:QEnLoss}) in 
a simple  approximate analytical  form. Using asymptotics of this  function
\begin{eqnarray}
\begin{aligned}
&\overline{H}(\chi)\approx {\frac {8\pi\sqrt {3}}{27}}{\chi}^{2},\quad  \chi\ll 1,  \\
&\overline{H}(\chi)\approx {\frac {32\pi\sqrt {3}}{243}}{2}^{2/3}\Gamma\!
\left(\frac{2}{3} \right) \chi^{2/3},\quad
\chi\gg 1.
\end{aligned}
\end{eqnarray}
we have found the following approximation
\begin{eqnarray}\label{eq:QLossApp}
\overline{H}(\chi)\approx\frac{8\pi\sqrt{3}}{27}\frac{\chi^2}{\left({\displaystyle 1+\frac{3}{4}\frac{(2\chi)^{2/3}}
{\sqrt{\Gamma\!\left(\frac{2}{3} \right) }}} \right)^2 }
\times \left(1+\frac{0.52\sqrt{\chi}(1+3\sqrt{\chi}-3.2\chi)}{1+0.3\sqrt{\chi}+17\chi+11\chi^2}\right)
\end{eqnarray}
The first two terms (before the sign $\times$) of Eq. (\ref{eq:QLossApp}) give right asymptotics at $\chi\ll 1$ and $\chi \gg 1$ and
provide an  accuracy better than $10\%$ for other values of $\chi$, whereas the inclusion of the last term in the brackets makes 
the  accuracy  better than $0.1\%$ for any  $\chi$.

The spectrum of the radiation in the quantum regime is expressed as
\begin{eqnarray}\label{eq:fq}
F_q(x,\tau)&=(1-\tau)F(x)+\tau^2 x K_{2/3}(x),
\quad x&=\frac{\tau}{1-\tau}\frac{E}{\epsilon_c},
\end{eqnarray}
where $\epsilon_c=\frac{3e\hbar B\sin\alpha}{2 mc}\gamma^2$ is the characteristic  energy of the emitted photon.
To use this function in Eq.~(\ref{eq:InRad}), the $\epsilon_c$ should be changed to $\hbar\omega_{*}$.
An  analytical approximation of this function can be obtained using the approximation for emissivity function of
the synchrotron radiation \citep{Aharonian2010}
\begin{eqnarray}
F(x)\approx 2.15 x^{1/3}(1+3.06 x)^{1/6}
\frac{1+0.884x^{2/3}+0.471 x^{4/3}}{1+1.64 x^{2/3}+0.974 x^{4/3}}e^{-x},
\end{eqnarray}
and
\begin{eqnarray}
x K_{2/3}(x)\approx 1.075 x^{1/3}(1+3.72 x)^{1/6}
\frac{1+1.58x^{2/3}+3.97 x^{4/3}}{1+1.53 x^{2/3}+4.25 x^{4/3}}e^{-x}.
\end{eqnarray}
Both approximations provide an accuracy better than $0.2\%$  for any 
value of the argument $x$.



\begin{thebibliography}{14}%
\makeatletter
\providecommand \@ifxundefined [1]{%
 \@ifx{#1\undefined}
}%
\providecommand \@ifnum [1]{%
 \ifnum #1\expandafter \@firstoftwo
 \else \expandafter \@secondoftwo
 \fi
}%
\providecommand \@ifx [1]{%
 \ifx #1\expandafter \@firstoftwo
 \else \expandafter \@secondoftwo
 \fi
}%
\providecommand \natexlab [1]{#1}%
\providecommand \enquote  [1]{``#1''}%
\providecommand \bibnamefont  [1]{#1}%
\providecommand \bibfnamefont [1]{#1}%
\providecommand \citenamefont [1]{#1}%
\providecommand \href@noop [0]{\@secondoftwo}%
\providecommand \href [0]{\begingroup \@sanitize@url \@href}%
\providecommand \@href[1]{\@@startlink{#1}\@@href}%
\providecommand \@@href[1]{\endgroup#1\@@endlink}%
\providecommand \@sanitize@url [0]{\catcode `\\12\catcode `\$12\catcode
  `\&12\catcode `\#12\catcode `\^12\catcode `\_12\catcode `\%12\relax}%
\providecommand \@@startlink[1]{}%
\providecommand \@@endlink[0]{}%
\providecommand \url  [0]{\begingroup\@sanitize@url \@url }%
\providecommand \@url [1]{\endgroup\@href {#1}{\urlprefix }}%
\providecommand \urlprefix  [0]{URL }%
\providecommand \Eprint [0]{\href }%
\providecommand \doibase [0]{http://dx.doi.org/}%
\providecommand \selectlanguage [0]{\@gobble}%
\providecommand \bibinfo  [0]{\@secondoftwo}%
\providecommand \bibfield  [0]{\@secondoftwo}%
\providecommand \translation [1]{[#1]}%
\providecommand \BibitemOpen [0]{}%
\providecommand \bibitemStop [0]{}%
\providecommand \bibitemNoStop [0]{.\EOS\space}%
\providecommand \EOS [0]{\spacefactor3000\relax}%
\providecommand \BibitemShut  [1]{\csname bibitem#1\endcsname}%
\let\auto@bib@innerbib\@empty
\bibitem [{\citenamefont {{Kelner}}\ and\ \citenamefont
  {{Aharonian}}(2012)}]{Kelner2012}%
  \BibitemOpen
  \bibfield  {author} {\bibinfo {author} {\bibfnamefont {S.}~\bibnamefont
  {{Kelner}}}\ and\ \bibinfo {author} {\bibfnamefont {F.}~\bibnamefont
  {{Aharonian}}},\ }\href@noop {} {\bibfield  {journal} {\bibinfo  {journal}
  {ArXiv e-prints}\ } (\bibinfo {year} {2012})},\ \Eprint
  {http://arxiv.org/abs/1207.6903} {arXiv:1207.6903 [astro-ph.HE]} \BibitemShut
  {NoStop}%
\bibitem [{\citenamefont {Alfven}\ and\ \citenamefont
  {Falthammar}(1963)}]{Alfven1963}%
  \BibitemOpen
  \bibfield  {author} {\bibinfo {author} {\bibfnamefont {H.}~\bibnamefont
  {Alfven}}\ and\ \bibinfo {author} {\bibfnamefont {C.~G.}\ \bibnamefont
  {Falthammar}},\ }\href@noop {} {\emph {\bibinfo {title} {{Cosmical
  Electrodynamics}}}}\ (\bibinfo  {publisher} {Oxford University Press},\
  \bibinfo {year} {1963})\BibitemShut {NoStop}%
\bibitem [{\citenamefont {{Landau}}\ and\ \citenamefont
  {{Lifshitz}}(1975)}]{Landau2}%
  \BibitemOpen
  \bibfield  {author} {\bibinfo {author} {\bibfnamefont {L.~D.}\ \bibnamefont
  {{Landau}}}\ and\ \bibinfo {author} {\bibfnamefont {E.~M.}\ \bibnamefont
  {{Lifshitz}}},\ }\href@noop {} {\emph {\bibinfo {title} {The Classical Theory
  of Fields}}}\ (\bibinfo  {publisher} {Butterworth-Heinemann},\ \bibinfo
  {address} {Oxford},\ \bibinfo {year} {1975})\BibitemShut {NoStop}%
\bibitem [{\citenamefont {{Cheng}}\ \emph {et~al.}(1986)\citenamefont
  {{Cheng}}, \citenamefont {{Ho}},\ and\ \citenamefont
  {{Ruderman}}}]{Cheng1986}%
  \BibitemOpen
  \bibfield  {author} {\bibinfo {author} {\bibfnamefont {K.~S.}\ \bibnamefont
  {{Cheng}}}, \bibinfo {author} {\bibfnamefont {C.}~\bibnamefont {{Ho}}}, \
  and\ \bibinfo {author} {\bibfnamefont {M.}~\bibnamefont {{Ruderman}}},\
  }\href {\doibase 10.1086/163829} {\bibfield  {journal} {\bibinfo  {journal}
  {\apj}\ }\textbf {\bibinfo {volume} {300}},\ \bibinfo {pages} {500} (\bibinfo
  {year} {1986})}\BibitemShut {NoStop}%
\bibitem [{\citenamefont {{Trubnikov}}(2000)}]{Trubnikov2000}%
  \BibitemOpen
  \bibfield  {author} {\bibinfo {author} {\bibfnamefont {B.~A.}\ \bibnamefont
  {{Trubnikov}}},\ }\href {\doibase 10.1134/1.1320080} {\bibfield  {journal}
  {\bibinfo  {journal} {Soviet Journal of Experimental and Theoretical
  Physics}\ }\textbf {\bibinfo {volume} {91}},\ \bibinfo {pages} {479}
  (\bibinfo {year} {2000})}\BibitemShut {NoStop}%
\bibitem [{\citenamefont {Schwinger}(1949)}]{Schwinger1949}%
  \BibitemOpen
  \bibfield  {author} {\bibinfo {author} {\bibfnamefont {J.}~\bibnamefont
  {Schwinger}},\ }\href {\doibase 10.1103/PhysRev.75.1912} {\bibfield
  {journal} {\bibinfo  {journal} {Phys. Rev.}\ }\textbf {\bibinfo {volume}
  {75}},\ \bibinfo {pages} {1912} (\bibinfo {year} {1949})}\BibitemShut
  {NoStop}%
\bibitem [{\citenamefont {{Berestetskii}}\ \emph {et~al.}(1982)\citenamefont
  {{Berestetskii}}, \citenamefont {{Lifshitz}},\ and\ \citenamefont
  {{Pitaevskii}}}]{Landau4}%
  \BibitemOpen
  \bibfield  {author} {\bibinfo {author} {\bibfnamefont {V.~B.}\ \bibnamefont
  {{Berestetskii}}}, \bibinfo {author} {\bibfnamefont {E.~M.}\ \bibnamefont
  {{Lifshitz}}}, \ and\ \bibinfo {author} {\bibfnamefont {L.~P.}\ \bibnamefont
  {{Pitaevskii}}},\ }\href@noop {} {\emph {\bibinfo {title} {Quantum
  Electrodynamics}}}\ (\bibinfo  {publisher} {Butterworth-Heinemann},\ \bibinfo
  {address} {Oxford},\ \bibinfo {year} {1982})\BibitemShut {NoStop}%
\bibitem [{\citenamefont {{Takata}}\ \emph {et~al.}(2004)\citenamefont
  {{Takata}}, \citenamefont {{Shibata}},\ and\ \citenamefont
  {{Hirotani}}}]{Takata2004}%
  \BibitemOpen
  \bibfield  {author} {\bibinfo {author} {\bibfnamefont {J.}~\bibnamefont
  {{Takata}}}, \bibinfo {author} {\bibfnamefont {S.}~\bibnamefont {{Shibata}}},
  \ and\ \bibinfo {author} {\bibfnamefont {K.}~\bibnamefont {{Hirotani}}},\
  }\href {\doibase 10.1111/j.1365-2966.2004.08270.x} {\bibfield  {journal}
  {\bibinfo  {journal} {\mnras}\ }\textbf {\bibinfo {volume} {354}},\ \bibinfo
  {pages} {1120} (\bibinfo {year} {2004})},\ \Eprint
  {http://arxiv.org/abs/arXiv:astro-ph/0408044} {arXiv:astro-ph/0408044}
  \BibitemShut {NoStop}%
\bibitem [{\citenamefont {{Hirotani}}(2008)}]{Hirotani2008}%
  \BibitemOpen
  \bibfield  {author} {\bibinfo {author} {\bibfnamefont {K.}~\bibnamefont
  {{Hirotani}}},\ }\href {\doibase 10.1086/595000} {\bibfield  {journal}
  {\bibinfo  {journal} {\apjl}\ }\textbf {\bibinfo {volume} {688}},\ \bibinfo
  {pages} {L25} (\bibinfo {year} {2008})},\ \Eprint
  {http://arxiv.org/abs/0810.0865} {arXiv:0810.0865} \BibitemShut {NoStop}%
\bibitem [{\citenamefont {{Levinson}}(2000)}]{Levinson2000}%
  \BibitemOpen
  \bibfield  {author} {\bibinfo {author} {\bibfnamefont {A.}~\bibnamefont
  {{Levinson}}},\ }\href {\doibase 10.1103/PhysRevLett.85.912} {\bibfield
  {journal} {\bibinfo  {journal} {Physical Review Letters}\ }\textbf {\bibinfo
  {volume} {85}},\ \bibinfo {pages} {912} (\bibinfo {year} {2000})}\BibitemShut
  {NoStop}%
\bibitem [{\citenamefont {{Levinson}}\ and\ \citenamefont
  {{Rieger}}(2011)}]{Levinson2011}%
  \BibitemOpen
  \bibfield  {author} {\bibinfo {author} {\bibfnamefont {A.}~\bibnamefont
  {{Levinson}}}\ and\ \bibinfo {author} {\bibfnamefont {F.}~\bibnamefont
  {{Rieger}}},\ }\href {\doibase 10.1088/0004-637X/730/2/123} {\bibfield
  {journal} {\bibinfo  {journal} {\apj}\ }\textbf {\bibinfo {volume} {730}},\
  \bibinfo {eid} {123} (\bibinfo {year} {2011})},\ \Eprint
  {http://arxiv.org/abs/1011.5319} {arXiv:1011.5319 [astro-ph.HE]} \BibitemShut
  {NoStop}%
\bibitem [{\citenamefont {{Aharonian}}\ \emph {et~al.}(2002)\citenamefont
  {{Aharonian}}, \citenamefont {{Belyanin}}, \citenamefont {{Derishev}},
  \citenamefont {{Kocharovsky}},\ and\ \citenamefont
  {{Kocharovsky}}}]{Aharonian2002}%
  \BibitemOpen
  \bibfield  {author} {\bibinfo {author} {\bibfnamefont {F.~A.}\ \bibnamefont
  {{Aharonian}}}, \bibinfo {author} {\bibfnamefont {A.~A.}\ \bibnamefont
  {{Belyanin}}}, \bibinfo {author} {\bibfnamefont {E.~V.}\ \bibnamefont
  {{Derishev}}}, \bibinfo {author} {\bibfnamefont {V.~V.}\ \bibnamefont
  {{Kocharovsky}}}, \ and\ \bibinfo {author} {\bibfnamefont {V.~V.}\
  \bibnamefont {{Kocharovsky}}},\ }\href {\doibase 10.1103/PhysRevD.66.023005}
  {\bibfield  {journal} {\bibinfo  {journal} {\prd}\ }\textbf {\bibinfo
  {volume} {66}},\ \bibinfo {eid} {023005} (\bibinfo {year} {2002})},\ \Eprint
  {http://arxiv.org/abs/arXiv:astro-ph/0202229} {arXiv:astro-ph/0202229}
  \BibitemShut {NoStop}%
\bibitem [{\citenamefont {{Bayer}}\ \emph {et~al.}(1973)\citenamefont
  {{Bayer}}, \citenamefont {{Katkov}},\ and\ \citenamefont {{Fadin}}}]{Bayer}%
  \BibitemOpen
  \bibfield  {author} {\bibinfo {author} {\bibfnamefont {V.~N.}\ \bibnamefont
  {{Bayer}}}, \bibinfo {author} {\bibfnamefont {V.~M.}\ \bibnamefont
  {{Katkov}}}, \ and\ \bibinfo {author} {\bibfnamefont {V.~S.}\ \bibnamefont
  {{Fadin}}},\ }\href@noop {} {\emph {\bibinfo {title} {Radiation of
  Relativistic Electrons}}}\ (\bibinfo  {publisher} {Atomizdat},\ \bibinfo
  {address} {Moscow},\ \bibinfo {year} {1973})\BibitemShut {NoStop}%
\bibitem [{\citenamefont {{Aharonian}}\ \emph {et~al.}(2010)\citenamefont
  {{Aharonian}}, \citenamefont {{Kelner}},\ and\ \citenamefont
  {{Prosekin}}}]{Aharonian2010}%
  \BibitemOpen
  \bibfield  {author} {\bibinfo {author} {\bibfnamefont {F.~A.}\ \bibnamefont
  {{Aharonian}}}, \bibinfo {author} {\bibfnamefont {S.~R.}\ \bibnamefont
  {{Kelner}}}, \ and\ \bibinfo {author} {\bibfnamefont {A.~Y.}\ \bibnamefont
  {{Prosekin}}},\ }\href {\doibase 10.1103/PhysRevD.82.043002} {\bibfield
  {journal} {\bibinfo  {journal} {\prd}\ }\textbf {\bibinfo {volume} {82}},\
  \bibinfo {eid} {043002} (\bibinfo {year} {2010})},\ \Eprint
  {http://arxiv.org/abs/1006.1045} {arXiv:1006.1045 [astro-ph.HE]} \BibitemShut
  {NoStop}%
\end{thebibliography}


%

\end{document}